\documentclass[journal=prl,showpacs,showkeys,twocolumn,notitlepage,floatfix]{revtex4-1}

\usepackage[scr=esstix,cal=boondox]{mathalfa} 
\usepackage{enumitem}
\usepackage{graphicx}
\usepackage{verbatim}
\usepackage{color}
\usepackage{amsmath,amsfonts,amsbsy,amssymb}
\usepackage{upgreek}

\def\L{\mathcal L}

\usepackage{rotating}

\def\Y{{\rm Y}}
\def\C{{\rm C}}

\begin{document}

\author{Paul N.\ Patrone}
\author{Lili Wang}
\author{Sheng Lin-Gibson}
\author{Anthony J.\ Kearsley}
\affiliation{National Institute of Standards and Technology \\ 100 Bureau Drive, Gaithersburg MD, 20899 USA}

\date{\today}
\title{Uncertainty Quantification of Antibody Measurements: Physical Principles and Implications for Standardization}

\begin{abstract}
Harmonizing serology measurements is critical for identifying reference materials that permit standardization and comparison of results across different diagnostic platforms.  However, the theoretical foundations of such tasks have yet to be fully explored in the context of  antibody thermodynamics and uncertainty quantification (UQ).  This has restricted the usefulness of standards currently deployed and limited the scope of materials considered as viable reference material.  To address these problems, we develop rigorous theories of antibody normalization and harmonization, as well as formulate a probabilistic framework for defining correlates of protection.  We begin by proposing a mathematical definition of harmonization equipped with structure needed to quantify uncertainty associated with the choice of standard, assay, etc.  We then show how a thermodynamic description of serology measurements (i) relates this structure to the Gibbs free-energy of antibody binding, and thereby (ii) induces a regression analysis that directly harmonizes measurements.  We supplement this with a novel, optimization-based normalization (not harmonization!) method that checks for consistency between reference and sample dilution curves.  Last, we relate these analyses to uncertainty propagation techniques to estimate correlates of protection.  {\it A key result of these analyses is that under physically reasonable conditions, the choice of reference material does not increase uncertainty associated with harmonization or correlates of protection.}  We provide examples and validate main ideas in the context of an interlab study that lays the foundation for using monoclonal antibodies as a reference for SARS-CoV-2 serology measurements.  
\end{abstract}

\maketitle

\section{Introduction}

The COVID-19 pandemic highlighted the importance of antibody tests as a means to characterize humoral response, e.g.\ in high-risk populations such as cancer patients \cite{Seronet}.  However, the rapid development of many different SARS-CoV-2 assays led to questions regarding the degree to which such measurements quantify immunity \cite{Correlates1,Correlates2,Correlates3,Correlates4}.  In response, public health and research institutions established reference materials to harmonize antibody scales \cite{Seronet,Seronet2,WHO1,WHO2}.  While these efforts facilitated quantitative comparison between assays, they also led to new questions: how do we compare reference materials, and how do we select the ``best'' one for a given clinical or research setting?\footnote{Certain commercial products are referenced (directly or indirectly) in this manuscript to clarify our theoretical analysis.  Such reference does not imply endorsement or approval of any kind by NIST.}  Moreover, the development of antibody standards did not suggest an obvious definition for correlates of protection, which remains an elusive concept to this day \cite{Correlates1,Correlates2,Correlates3,Correlates4,Binding}.

The difficulty of addressing these questions arises from a fundamental ambiguity in the information one typically seeks from serology measurements.  In particular, binding assays quantify the relative number of antibodies in a sample that attach to a substrate.  However, the corresponding binding kinetics  depend on details of the assay itself, such as the choice of substrate epitopes \cite{Affinity}.  Thus the concept of an absolute and independent ``bound antibody number'' is not meaningful without further information about the underlying chemical reaction.\footnote{This mirrors the challenge of estimating binding kinetics for emerging SARS-CoV-2 variants.}  This ambiguity becomes {\it worse} when we recognize that all reference materials suffer the same problem, which leads to the possibility of expressing an ill-defined sample concentration in terms similarly ill-defined standard.  
These conceptual issues are compounded by the fact that typical single-point reference calibrations, e.g.\ based on end-point titers \cite{Endpoint}, suffer from extrapolation error \cite{singlepoint}.  Thus, meaningful comparison of antibody standards cannot be realized without identifying and quantifying the sources of uncertainty that affect their use.

The present work addresses these problems via a hierarchy of data analyses and statistical methods that leverage tools from thermodynamics and conditional probability.  We first motivate this hierarchy through a Gibbs free-energy description of antibody binding, which (surprisingly) implies that only the assays, not the reference, control the degree to which harmonization is possible.  Importantly, this theory induces a probabilistic model that can be used to check the thermodynamic assumptions by quantifying -- and in some cases removing -- measurement variability due to: (i) choice of reference material; (ii) assay; (iii) instrument and operator effects; (iv) uncertainty inherent in samples; and (v) interactions between these elements.  The first and last of these are then used to determine if and by how much a specific reference material increases uncertainty in harmonization, which becomes a metric for comparing standards.  We also couple this framework to neutralization measurements to quantify confidence that an individual has a certain degree of protection (e.g.\ against SARS-CoV-2 infection) based only on binding assay results.  This permits us to determine whether a given reference also impacts correlates of protection, not just harmonization.  Throughout, we validate these ideas in the context of an interlab study that establishes a reference scale in terms of synthetic, monoclonal antibodies (mAbs) \cite{InterlabLili}.

A key theme that permeates this work is the need to incorporate uncertainty quantification (UQ) into all aspects of the data analysis.  While this entails obvious tasks such as statistical modeling and uncertainty propagation \cite{SmithUQ}, we take a much broader perspective that UQ should also include analyses that assess and increase confidence in the measurement and harmonization processes {\it per se}.  For example, by virtue of being both monoclonal and synthetic, the kinetic properties of mAbs are inherently suspect for purposes of harmonization.  Thus we develop a novel normalization procedure that tests the physical consistency of each dilution series with respect to that of the reference material, validating that mAbs behave like human-derived samples.  Moreover, our probabilistic models are largely {\it derived} from known thermodynamic properties of antibodies to ensure that harmonization is based on rigorous physical principles, not simply heuristic justifications.

A main challenge in adopting this comprehensive approach to UQ is the need to revisit and clarify  seemingly elementary and resolved ideas in serology.  For example, the concept of harmonization has been used in a variety of contexts \cite{harmonization1,harmonization2,NCI}, but to the best of our knowledge, it has not been defined with the precision needed to fully ground it in metrology.  Thus, a preliminary task in our analysis is to mathematically define this concept and equip it with the structure needed to permit later UQ.  {\it The implications of this exercise are not trivial.  It uncovers hidden structure in the definition of harmonization that has yet to be exploited and provides a direct connection to the physics of antibody measurements.  Moreover, it reveals that harmonization and normalization are not the same task.}  In a similar vein, we propose a precise mathematical construction that can be used in a decision-theory framework to define correlates of protection.  This exercise is likewise useful for quantifying the corresponding impact of a given reference material.

This work is motivated by the fact that current methods to assess fitness of purpose for serology standards are insufficiently grounded in UQ and focus primarily on their {\it technical performance,} sometimes ignoring issues such as ease of manufacturing, distribution times, etc.  For example, there is widespread belief that harmonization via SARS-CoV-2 standards can only be achieved using human-derived, pooled references, although to the best of our knowledge there are no studies validating this conjecture.  As a result, the development of current SARS-CoV-2 standards (which are all human derived) took nearly a year \cite{WHO1,WHO2} despite being needed much sooner.  Moreover,  such standards are inherently limited stock and must contend with changes between lots \cite{WHO3}.  These issues suggest the need to better understand the impacts of using alternatives such as mAbs, which permit better scale-up and quicker development.  {\it Indeed our companion manuscript finds -- and the present work justifies -- that mAbs and human-derived standards are identical from a performance standpoint when using our physically derived definition of harmonization.}  That this conclusion arises from a novel, thermodynamics-based UQ analysis emphasizes the central, and heretofore missing role of uncertainty in the selection of standards.

A limitation of our work is that we do not define or fully connect our work with a real-world definitions of immunity.  This is due primarily to lack of available data \cite{COVIDUQ1,COVIDUQ2}.  Studies that address this problem would need to connect information about neutralization measurements to notions of risk and clinical outcomes.  To the best of our knowledge, the these latter tasks remain open challenges in the SARS-CoV-2 testing community and thus fall beyond the scope of our work \cite{COVIDUQ1,COVIDUQ2}.  However, in Secs.\ \ref{sec:definitions} and \ref{sec:neuts} we provide tools that may help to address such challenges.

The rest of the manuscript is organized as follows.  In Sec.\ \ref{sec:notation}, we provide physical and mathematical definitions of key concepts used throughout the manuscript.  Section \ref{sec:perspectives} gives a global overview of our analysis hierarchy by deriving a thermodynamic theory of antibody binding that induces a probabilistic perspective needed for harmonization.  The remaining sections fully develop the elements of this hierarchy.  In particular, the insight gained from the thermodynamic model and definitions leads to a new technique for antibody normalization in Sec.\ \ref{sec:affine}.  Section \ref{sec:betweenuc} then shows how to combine this result with the probabilistic model of Sec.\ \ref{sec:perspectives} to quantify the various sources of uncertainty (including the reference material) associated with binding assays.  Section \ref{sec:neuts} uses these results to estimate neutralization titers and uncertainty therein based only on binding measurements.  Section \ref{sec:discussion} provides historical context for our work, discusses limitations, and considers future directions.


\section{Notation and Terminology}
\label{sec:notation}

Biometrology, to which antibody testing pertains, is an inherently interdisciplinary field.  We maintain that mathematical rigor is just as important as experimental protocol, since the former informs how to interpret data obtained from the latter.  As a result, this manuscript does not shy away from theory, although this is not meant to discourage any particular audience.  Rather, we take the perspective that a convergence of language across disciplines is necessary to realize the full potential of biometrology.  Thus, at a minimum, we aim to make the core ideas accessible and intuitive to any reader with a modest background in serology testing, with full details accessible to those wishing to verify the underlying mathematics.     To these ends, certain definitions are fundamental to our discussion.  

\subsection{Definitions}
\label{sec:definitions}

\begin{itemize}
\item[I.] A concentration $\hat c$ is {\bf normalized} if it is given in units of antibodies per volume.     
\item[II.] The word {\bf sample} always refers a specimen taken from a human and to which a serology test is applied.
\item[III.] The words {\bf reference} and {\bf standard} always refer to a measurand, which can be synthetic or human-derived, used to normalize measurements taken on samples.  
\item[IV.]  In physical terms, we interpret {\it harmonization} as the process of determining a mathematical rule that tells one how to modify the numerical value of normalized antibody concentrations for each assay so that their corresponding measurements all agree and can be used interchangeably.\footnote{We do not assume that normalization and harmonization are necessarily identical.  In the next section, however, we determine conditions under which these tasks are the same.} {\it This rule is only considered meaningful if it does not depend on the sample}, but only the assay, reference, and concentration values.  In more mathematical terms, let $s=1,2,...,S$ and $n=1,2,...,N$ index samples and assays for some maximum values $S$ and $N$.  Also let $r$ denote a fixed reference.  We say that the assays are {\bf harmonized} by reference $r$ if for any normalized concentrations $\hat c_{s,n,r}$ and $\hat c_{s,n',r}$ (corresponding to sample $s$  measured with assays $n$ and $n'$ and normalized by $r$), we can find a function $T(n,c,r)$ such that 
\begin{align}
T(n,\hat c_{s,n,r},r)=T(n',\hat c_{s,n',r},r) = {\chi}_{s,r}, \label{eq:harmonization_def}
\end{align}
where ${\chi}_{s,r}$ is an assay-independent {\bf consensus value} associated with sample $s$.  {\it Consistent with the above physical intuition, this function does not depend directly on the sample index $s$, only its normalized concentrations $\hat c_{s,n,r}$.}    See Refs.\ \cite{harmonization1,harmonization2} and the references therein for related ideas.  
\item[V.] If we identify parameters $\epsilon_{s,n}$ (which could be random) such that 
\begin{align}
T(n,\hat c_{s,n,r},r)(1 + \epsilon_{s,n})&=T(n',\hat c_{s,n',r})(1+ \epsilon_{s,n'}) \nonumber \\
&= {\chi}_{s,r}, \label{eq:approx_harmonization_def}
\end{align}
we say that the assays can be {\bf approximately harmonized} with a relative uncertainty quantified by the $\epsilon$. In principle we could let $\epsilon_{s,n}$ also depend on $r$, but in later sections we find this assumption unnecessary.   In a slight abuse of terminology, we sometimes refer to the concentrations $T(n,\hat c_{s,n,r},r)$ as harmonized (without the modifier ``approximately'')  when the meaning is clear from context.  
\item[VI.] Let $X$ and $Z$ denote the outputs of two distinct types of measurements (e.g.\ a binding and neutralizing assay), and let $\cal P$ be a percentage satisfying $0 < \cal P \le 1$.  We say that $X^\star$ is the $\cal P${\bf -correlate level} for $Z^\star$ if it is the smallest value such that $X \ge X^\star$ implies that the probability of $Z \ge Z^\star$ is greater than $\cal P$.  Mathematically, 
\begin{align}
X^\star \!=\! \mathop{\rm min}_{\mathcal X} \left\{{\mathcal X}: X \ge {\mathcal X} \! \implies \! {\rm P}(Z \ge Z^\star) \!\ge\! \cal P \right\}, \label{eq:preCOP}
\end{align}
where ${\rm P}(Z \ge Z^\star)$ is the probability that $Z \ge Z^\star$.  While seemingly abstract, this definition is fundamental to our notion of {\bf correlates of protection}.  In particular, Eq.\ \eqref{eq:preCOP} enables statements of the form, ``A binding level of at least $X^\star$ implies that a neutralization level is at least $Z^\star$ with a probability of 95\% or greater.''  Note that $X^\star$ is a function of $Z^\star$ and the probability $\cal P$.  
\end{itemize}

\subsection{Notation and Conventions}

\begin{itemize}
\item We treat Gibbs free-energies $G$ (and differences $\Delta G$ thereof) as dimensionless, having been divided by the temperature expressed in units of energy.  
\item We denote a normal random variable with mean $\mu$ and variance $\sigma^2$ via $\mathcal N(\mu,\sigma^2).$
\item For clarity, we reserve certain indices for special purposes.  Lowercase $m$ refers to either a sample or a reference.  Lowercase $s$ and $r$  refer exclusively to samples and references.  Lowercase $n$ always refers to an assay.  Lowercase $k$ is used generically as an integer index except in any of the previously mentioned cases.  {\it We caution the reader that throughout the manuscript, indices are often the primary (and sometimes only) way we indicate dependence between variables.}
\item In certain cases, we need to non-dimensionalize the arguments of transcendental functions by dividing through by the units.  In such cases, we use the symbol $U_\star$ to indicate a quantity whose value is 1 multiplied by the units associated with $\star$. 
\end{itemize}

\subsection{Units Associated with Antibody Normalization}
\label{subsec:units}

Normalization aims to quantify the concentration $c$ of antibodies that bind to a substrate.  However, this concentration is never measured directly; neither is it possible to measure the {\it total concentration} of antibodies except as part of the manufacturing process for certain reference materials.  Instead, the instrument outputs some numerical value such as median fluorescence intensity (MFI) $F$ expressed, for example, in units of voltage.  This $F$ is {\it typically} assumed to be proportional to the bound concentration.\footnote{In later sections we find it more convenient to dispense with this linearity assumption.}  The constant of proportionality, which we denote $\Gamma$, has units such as voltage (i.e.\ some proxy for MFI) times volume per number of bound antibodies; viz.\ 
\begin{align}
F=\Gamma c.
\end{align}
In turn, $c$ is assumed to be proportional to the total concentration $y$ of antibodies of a fixed type via the theory described in Sec.\ \ref{subsec:thermo}; see also Refs.\ \cite{Endpoint,singlepoint}.  We denote the corresponding proportionality constant by $K$, which is dimensionless but should be thought of as having units of bound antibody number per total antibody number.  Thus, one trivially finds
\begin{align}
c=Ky \implies F= \Gamma Ky,
\end{align}
where the product $\Gamma K$ has units of voltage per concentration of total antibodies.  

Normalization is typically performed by computing a ratio of the form
\begin{subequations}
\begin{align}
\hat y_{s,r} &= \frac{F_s}{F_r} \label{eq:old_norm}\\
& = \frac{c_s}{c_r} = \frac{K_s y_s}{K_r y_r} \label{eq:our_norm}
\end{align}
\end{subequations}
where subscripts $s$ and $r$ denote corresponding quantities for a sample and reference.  While this $\hat y_{s,r}$ is ostensibly a normalized antibody value, it is dimensionless; i.e.\ it is the number of bound sample antibodies per bound reference antibody.  To make $\hat y_{s,r}$ consistent with Definition I, we multiply by through $y_r$ (which is assumed to be known), yielding
\begin{align}
\hat c_{s,r} = \hat y_{s,r} y_r. \label{eq:renorm}
\end{align}
The left-hand side (LHS) of Equation \eqref{eq:renorm} has units of (total) antibodies per volume.

Four comments are in order.  

First, recognize that many of the above quantities ($\hat c_{s,r}$, $\hat y_{s,r}$, $c_r$, $c_s$, etc.) depend on the choice of assay.  In subsequent sections we make this explicit by including the subscript $n$.  Here, we have suppressed the assay dependence since it is not central to a discussion of units.

Second, a more appropriate definition of $\hat c_{s,r}$ would be $\hat y_{s,r} c_r$, since this has the biologically relevant units of bound antibodies per volume \cite{WHO1,WHO2,NCI}.\footnote{When $y_r$ is unknown, as is the case for human-derived, pooled samples, its value is assigned arbitrarily \cite{WHO1,WHO2,NCI}.  In this case, it is equally valid to fix $c_r$ instead.  This is purely a semantic choice that does not play a role in our analysis.  }  However, $K_r$ is generally unknown, which implies the same for $c_r$.  Thus, one can only normalize bound antibody concentration up to an unknown scale factor associated with the $K_r$, which turns out to be an equilibrium constant; see Sec.\ \ref{subsec:thermo}.

Third, while $\hat y_{s,r}$ and $\hat c_{s,r}$ are nominally different, they are interchangeable from a theoretical standpoint because of Eq.\ \eqref{eq:renorm}.  The quantity $\hat y_{s,r}$ is mathematically more convenient because it is dimensionless, and we often use this form of the normalized measurement.  We refer to $\hat y_{s,r}$ as a {\it scaled} antibody concentration and take ${\cal y}_{s,r}$ to be the corresponding consensus value.  

Fourth, normalization is typically accomplished by computing the ratio given by Eq.\ \eqref{eq:old_norm}, i.e.\ directly in terms of the measurement outputs.  However, the equivalence of Eqs.\ \eqref{eq:old_norm} and \eqref{eq:our_norm} foreshadows that directly accessing the dimensionless ratio $c_s/c_r$ is another route to estimating $\hat y_{s,r}$.  In Sec.\ \ref{sec:affine} we show that this is preferable.


\section{Global Perspectives on our Analysis} 
\label{sec:perspectives}

\subsection{Thermodynamic Perspective}
\label{subsec:thermo}

A binding antibody assay can be viewed as a chemical reaction \cite{Affinity,massaction}.  A free antibody Y attaches to a substrate B to create a bound pair YB; viz.
\begin{align}
{\rm Y} +  {\rm B} \longleftrightarrow {\rm YB} = \C, \label{eq:reaction}
\end{align}
where $\C$ stands for the antibody-substrate complex.  This reaction is assumed to be reversible, so that the system reaches an equilibrium described by detailed balance \cite{massaction,Affinity2}.  For a fixed antibody type $\Y_m$ (associated with the $m$th sample or reference) and substrate ${\rm B}_n$ associated with an the $n$th assay or antigen epitope, the equilibrium constant $\hat K$ can depend on {\it both} the antibody and the substrate.  That is, 
\begin{align}
\frac{c_{m,n}}{(y_m-c_{m,n}) (b_n-c_{m,n})} = \hat K_{m,n}, \label{eq:equilibrium}
\end{align}
where $c_{m,n}$, $y_m$, and $b_n$ are the concentrations of $\C_{m,n}$, all antibodies (free and bound), and all substrates ${\rm B}_n$.  The $n$-dependence of $\hat K_{m,n}$ reflects the fact that changing the substrate can alter the equilibrium concentration, and hence number of bound antibodies.  The physical intuition for this dependence arises from the definition
\begin{align}
\frac{\hat K_{m,n}}{U_{\hat K}} = e^{-\Delta G_{m,n}} \to \Delta G_{m,n} = -\ln (\hat K_{m,n}/U_{\hat K}),\label{eq:Gibbs}
\end{align}
where $\Delta G_{m,n}$ is the Gibbs free-energy change associated with Eq.\ \eqref{eq:reaction} \cite{Pathria}.  That is, changing the substrate and/or antibody alters the free-energy landscape, and thus the equilibrium constant.  Note that Eq.\ \eqref{eq:equilibrium} only models antibody affinity, but not {\it avidity} (i.e.\ capacity for multivalent binding); see Ref.\ \cite{Monovalent} for justification in the context of the examples considered herein.


From a measurement standpoint, it is desirable for $c_{m,n}$ to be independent of the substrate concentration; doing so ensures that the former increases linearly with the total antibody concentration \cite{linearity}.  A straightforward Tayor expansion of Eq.\ \eqref{eq:equilibrium} reveals that this condition is approximately satisfied if either (i) $b_n \gg y_m$ and $\hat K_{m,n} \gtrsim \mathcal O(1)$  unit volume, or (ii) $\hat K_{m,n} \ll 1$ unit volume for $b_n$ and $y_m$ order one concentrations.  We henceforth assume that either (i) or (ii) is true,\footnote{It is necessary to distinguish nonlinearity in the measurement due to Eq.\ \eqref{eq:equilibrium} from nonlinear effects due to detection equipment such as photodetectors.  This distinction is important for the analysis in Sec.\ \ref{sec:affine}.} which yields the approximate model
\begin{align}
c_{m,n} = K_{m,n} y_m, \label{eq:lineareq}
\end{align}  
where $K_{m,n}$ is an appropriately rescaled equilibrium constant.\footnote{\textcolor{black}{When $\hat K_{m,n} \ll 1$  unit volume, $K_{m,n}=\hat K_{m,n}b_n$.  When $b_n \gg y_m$, one finds $K_{m,n}=1$.}}  

An interesting question that motivates our harmonization analysis is whether the equilibrium constant is {\it separable}, meaning it can be expressed in the form 
\begin{align}
K_{m,n}=\kappa_{\Y_m} \kappa_{{\rm B}_n}, \label{eq:separability}
\end{align}
for some constants $\kappa_{\Y_m}$ and $\kappa_{{\rm B}_n}$ depending only on the sample or assay, but not both.  Physically, Eq.\ \eqref{eq:separability} is interpreted as the condition in which the antibodies' contribution to the equilibrium constant is independent of the substrate contribution.\footnote{Note that the $\kappa_{\Y_m}$ and $\kappa_{{\rm B}_n}$ are not determined uniquely by Eq.\ \eqref{eq:separability}.  We can always define new constants $\hat \kappa_{\Y_m}= \kappa_{\Y_m}/\alpha$ and $\hat \kappa_{{\rm B}_n}=\kappa_{{\rm B}_n}\alpha$ for any positive constant $\alpha$ such that the product  $K_{m,n}=\hat \kappa_{\Y_m}\hat \kappa_{{\rm B}_n} = \kappa_{\Y_m} \kappa_{{\rm B}_n}$ is unchanged.}  In the context of Eq.\ \eqref{eq:Gibbs}, this implies that the change in free energy can be expressed as a sum
\begin{align}
\Delta G_{m,n} = \Delta G_m + \Delta G_n, \label{eq:Gibbs_separable}
\end{align}
where, in a slight abuse of notation, we treat $\Delta G_m$ and $\Delta G_n$ as distinct quantities depending on either the sample or substrate, but not both.  Heuristically Eq.\ \eqref{eq:Gibbs_separable} is plausible if {\it any antibody} that binds to a fixed antigen (corresponding to constant $n$)  always changes the latter's conformation in the same way, so that $\Delta G_{m,n}$ only varies with the internal energy and entropy differences between the antibodies themselves.

To understand the usefulness of separability, recall that antibody normalization amounts to determining the ratio $c_{s,n}/c_{r,n}$ for sample $s$, reference $r$, and a fixed assay $n$.  Separability then amounts to the condition that
\begin{align}
\frac{c_{s,n}}{c_{r,n}} &= \frac{\kappa_{\Y_s}y_s}{\kappa_{\Y_{r}}y_{r}} \label{eq:ideal_harmonization} 
\end{align}
which is equivalent to
\begin{align}
\ln\left(\frac{c_{s,n}}{c_{r,n}} \right) = \ln \left(\frac{y_s}{y_{r}} \right) - \Delta G_s + \Delta G_r. \label{eq:logsep}
\end{align}
{\it In other words, relative concentrations of bound antibodies are independent of the assay being used for the measurement}, since the right-hand side (RHS) has no dependence on $n$.  Separability therefore implies that normalization automatically harmonizes assays in the sense of Definition I, and we can simply set the function $T(n,c)=c$.

If we relax the separability assumption, harmonization is no longer guaranteed.\footnote{We do not explore the theoretical question of whether separability is  {\it necessary} for harmonization.
}
To see this, assume that 
\begin{align}
K_{m,n} = \kappa_{\Y_m} \kappa_{{\rm B}_n}\exp (-\Delta g_{m,n}), \label{eq:separability_approx}
\end{align}
where $\Delta g_{m,n}$ is a relative free-energy deviation from separability.  We require that $\Delta g_{m,n}$ depend non-trivially on both of its indices.\footnote{\textcolor{black}{More precisely, when viewed as a matrix with elements $s,n$, we require $K_{s,n}$ to have rank greater than one.}}  Taking the logarithm of $c_{s,n}/c_{r,n}$ yields
\begin{align}
\ln \left( \frac{c_{s,n}}{c_{r,n}} \right ) &= \ln \left( \frac{\kappa_{\Y_s}y_s}{\kappa_{\Y_{r}}y_{r}} \right )  -\Delta g_{s,n} + \Delta g_{r,n}. \label{eq:linear_disharmony}
\end{align}
The term $\Delta g_{s,n}$ is problematic; it implies that the normalized concentration depends on the free-energy of the specific sample-assay pair, which is nominally inconsistent with harmonization.

In practice, we expect that Eq.\ \eqref{eq:linear_disharmony} is a more realistic description of antibody measurements; biological variation between human samples and differences in substrate epitopes will cause some antibodies to bind differently to some assays, invalidating the heuristic picture described below Eq.\ \eqref{eq:Gibbs_separable}.  Thus it is not clear that exact harmonization is possible.  However, we can still recover the weaker notion of approximate harmonization given by Def. V.  In particular, were it possible to determine the $\Delta g_{r,n}$, one could define a transformation
\begin{align}
T(n,\hat c,r)= \hat c \exp(-\Delta g_{r,n}), \label{eq:bias_harmonization}
\end{align}
where we now reveal the explicit dependence of $T$ on the reference material.  Combined with Eq.\ \eqref{eq:linear_disharmony}, this would imply that
\begin{align}
T(n,\hat c_{s,n,r},r)(1\!+\!\epsilon_{s,n}) &= T(n',\hat c_{s,n',r},r)(1\!+\!\epsilon_{s,n'}) \nonumber \\ 
& = {\chi}_{s,r} \label{eq:appox_harmonization_bias} \\
\epsilon_{s,n} &= \exp(\Delta g_{s,n}) - 1
\end{align}
yields a consensus value.  

A notable conclusion of Eq.\ \eqref{eq:linear_disharmony} is that lack of harmonization has nothing to do with the choice of reference material.  It is due solely sample-assay dependent effects $\epsilon_{s,n}$, since negating these implies that Eq.\ \eqref{eq:bias_harmonization} is an exact harmonization rule according to Definition V.  Physically, this conclusion arises from the simple fact that all samples are normalized by the same reference material, so they share a common bias associated with $\Delta g_{r,n}$.  Our companion manuscript validates this result for a collection of several monoclonal antibodies \cite{InterlabLili}.  The next section motivates a UQ analysis to test the validity of Eq.\ \eqref{eq:appox_harmonization_bias}.

\subsection{Induced Probabilistic Perspective}

Equation \eqref{eq:appox_harmonization_bias} begs two questions: (i) how do we validate the underlying model; (ii) how do we use it to harmonize assays?  In addition, we wish to connect binding measurements to neutralization assays to derive a more rigorous definition of correlates of protection.  Our main goal in this section is to show how Eq.\ \eqref{eq:appox_harmonization_bias} induces a  probabilistic framework that addresses these questions.

First note that the quantities $\Delta g_{s,n}$ and $\Delta g_{r,n}$ are in general unknown, since it is unreasonable to perform detailed measurements characterizing equilibrium constants for all samples and assays.  For a fixed assay, however, $\Delta g_{r,n}$ is common to all samples, whereas $\Delta g_{s,n}$ is sample dependent.  Rearranging Eq.\ \eqref{eq:linear_disharmony} implies that
\begin{align}
\ln(\hat y_{s,n,r}) - \Delta g_{r,n} +  \Delta g_{s,n} = \ln({\cal y}_{s,r}) \label{eq:regression_model}
\end{align}
which suggests interpreting $\Delta g_{r,n}$ as a constant, reference-dependent {\it bias} and $\Delta g_{s,n}$ as a sample-assay-dependent realization of a random variable.  Using Eq.\ \eqref{eq:regression_model} as the basis for a regression analysis applied to a collection of data should therefore yield estimates of these quantities.  {\it Doing so yields all of the information needed to construct $T$ and $\epsilon$ in Eq.\ \eqref{eq:appox_harmonization_bias}, thereby harmonizing the assays according to Def.\ VI.}

We can validate the model by testing the degree to which the $\Delta g_{s,n}$ empirically depend on the reference.  Despite the assumption that these quantities are random, Eq.\ \eqref{eq:regression_model} indicates that remain they {\it point-wise constant} with changing $r$, provided the $\Delta g_{r,n}$ are correctly determined.  This restrictive criterion implies that the residuals between the bias-corrected and consensus antibody estimates should be invariant to the reference, which can be directly checked.  We pursue this and related ideas in Sec.\ \ref{sec:betweenuc}.

In order for this analysis to be meaningful, however, it is necessary to account for uncertainty inherent in the measurement process.  The true $\hat y_{s,n,r}$ are never known exactly due to effects such as pipetting error, instrument artifacts, etc.  Thus, it is important to ensure that the associated measurement variability is not confused with the $\Delta g_{r,n}$ and $\Delta g_{s,n}$.  We therefore postulate that the quantity $\bar y_{s,n,r}$ one measures is related to $\hat y_{s,n,r}$ via the equation
\begin{align}
\ln (\hat y_{s,n,r}) = \ln(\bar y_{s,n,r}) + \delta_{s,n,r} \label{eq:withinuc}
\end{align}
where $\delta_{s,n,r}$ is a  ``within-lab'' uncertainty that models the effects described immediately above; see Ref.\ \cite{Rukhin}.  The dependence of $\delta$ on $s$, $n$, and $r$ is largely incidental, since this quantity should be estimated separately for each such triple.  The $\delta$ does not account for assay and reference-specific effects.  Combining Eqs.\ \eqref{eq:regression_model} and \eqref{eq:withinuc} then yields
\begin{align}
\ln(\bar y_{s,n,r}) - \Delta g_{r,n} + \Delta g_{s,n} + \delta_{s,n,r} = \ln({\cal y}_{s,r}). \label{eq:fullmodel}
\end{align}

In practice, the values of $\bar y_{s,n,r}$ and $\delta_{s,n,r}$ are given by direct measurement outputs, while the remaining quantities must be determined from the data via regression; such details are postponed until Sec.\ \ref{sec:betweenuc}.  However, several comments are in order.

First, while $\delta_{s,n,r}$ nominally depends on all three of its indices, to extremely good approximation it exhibits separability of the form 
\begin{align}
\delta_{s,n,r}= \delta_{s,n} + \delta_{r,n}. \label{eq:separablenoise}
\end{align}
This arises from the fact that normalized measurements are computed as a ratio of concentrations.  Thus logarithms thereof separate uncertainty in a way that mirrors the terms $\Delta g_{s.n}$ and $\Delta g_{r,n}$ in Eq.\ \eqref{eq:regression_model}.  Moreover, the quantities $\delta_{r,n}$ and $\Delta g_{r,n}$ act on normalized samples in the same way, so that a regression analysis essentially absorbs the former into the latter.  Thus, while we keep track of the $r$ dependence of $\delta_{s,n,r}$ throughout, only $\delta_{s,n}$ manifests directly at the end.

Second, observe that Eq.\ \eqref{eq:fullmodel} yields a probability density for a consensus value given a sample measurement from a specific assay scaled with reference $r$.  We denote this function by $P({{\cal y}_{s,r}}|\bar y_{s,n,r},r,n)$, meaning that the probability of a specific ${\cal y}_{s,r}$ is conditioned on the triple $(\bar y_{s,n,r},r,n)$.  If we likewise construct a probability $P(\nu|\cal y)$ of a neutralizing value $\nu$ (e.g.\ expressed as a 50 \% neutralizing titer or NT50) conditioned on the consensus value, then the probability of a neutralizing value given a scaled binding value is \cite{RW}
\begin{align}
\mathcal P(\nu|\bar y_{s,n,r},r,n) = \int {\rm d}{\cal y} \,\, P(\nu|{\cal y})P({\cal y}|\bar y_{s,n,r},r,n). \label{eq:neutprob}
\end{align}
Equation \eqref{eq:neutprob} provides actionable information.  Defining $\nu_t$ as a lower neutralizing threshold that guarantees a degree of immunity, one can quantify the probability $\mathcal P(\nu \ge \nu_t)$ that a person is protected by computing the integral
\begin{align}
\mathcal P(\nu \ge \nu_t | \bar y_{s,n,r},r,n) = \int_{\nu_t}^\infty {\rm d}\nu \,\, \mathcal P(\nu|\bar y_{s,n,r},r,n) \label{eq:proprob}
\end{align}
One can then find the minimum measured binding level $\bar y_{\min}$ that guarantees the corresponding $\nu$ is above $\nu_t$ with confidence $\mathcal P(\nu \ge \nu_t) > 95\%$, for example.  In this case, $\bar y_{\min}$ would be the 95\% correlate level for $\nu_t$; see Definition VI.


\section{Affine Transformations for Dilutions Series: Global Method for Antibody Normalization}
\label{sec:affine}

In this section, we address the task of estimating the scaled antibody level $\hat y_{s,n,r}$.  Traditional methods rely on identifying a linear region of the detector and diluting samples so that their measurement values fall within that range \cite{Endpoint,linearity}.  However, this approach discards the majority of available information if multiple dilutions are measured, as is often the case.  Moreover, the normalized antibody estimates are single-point calibrations, which may not be stable against measurement uncertainty \cite{singlepoint}.  We propose an alternate method that uses the entire dilution series while generalizing the theory of Ref.\ \cite{singlepoint}.

Without loss of generality, we assume that the measurement value $F$ corresponds to MFI.  Let $x$ denote the number (not concentration!) of bound antibodies in a sample, and assume that there exists a range $[x_{\min},x_{\max}]$ over which $F(x)$ is on average a strictly monotone increasing function of $x$.  The physical interpretation is straightforward: more antibodies yield more fluorescence.  We supplement this with this assumption that on average, $F(x)$ approaches lower and upper limits $F_{\min}$ and $F_{\max}$ (which we can always take to be positive) as $x\to 0$ or $x\to \infty$.  Physically $F_{\min}$ and $F_{\max}$ can be interpreted as a noise-floor and detector saturation threshold.  It is important to distinguish these sources of nonlinearity, which are instrument artifacts, from effects associated with nonlinear dependence of $c_{s,n}$ on $y_s$ as expressed by Eq.\ \eqref{eq:equilibrium}.  We always assume that $c$ (and thus $x$) is linear in $y$, even when fluorescence $F$ is nonlinear in $c$.

\begin{figure}
\includegraphics[width=8cm]{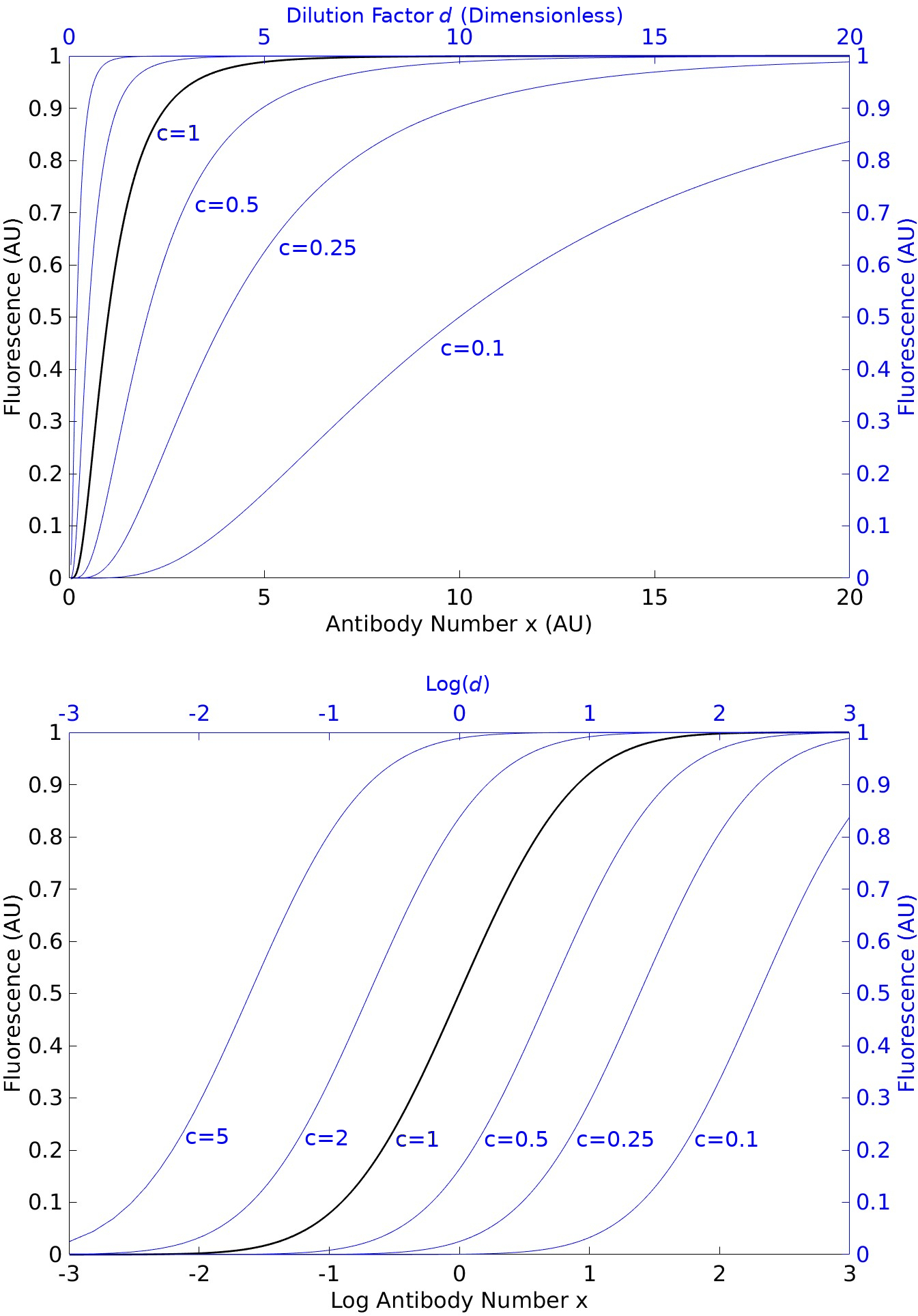}\caption{Plots of synthetic data motivating the algorithm associated with Eq.\ \eqref{eq:equivalence}.  Note carefully that the colors are matched to axes and do not have interchangeable interpretations.  Our normalization algorithm prescribes a mathematical method for reconciling their differences.  In both plots, $d$ corresponds to a dilution factor; see Eq.\ \eqref{eq:forward} for details.  In the top plot, the left-most blue curves correspond to $c=5$ and $c=2$ from left to right.   In all plots, the black dilution curve corresponds to a reference material.   Changing the number of antibodies in the reference traces out the dilution curve.  Equivalently, if we fix the concentration (say to $c=1$), diluting ($d < 1$) or concentrating ($d>1$) the reference yields the same curve.  {\it Importantly, this dual interpretation only applies to the reference.}  Taking a sample for which $c > 1$ implies that the sample must be diluted {\it more} relative to the reference to generate the same fluorescence signal.  In order to collapse the blue curve onto the reference, we must scale the dilution by a factor of $c$ (i.e.\ $c_s/c_r$ for $c_r=1$ and $c_s=c$), which corresponds to a horizontal shift on a log scale (bottom plot).  A similar interpretation applies to the case $c<1$.   }\label{fig:motivation}
\end{figure}

Because the function $F(x)$ only depends on the {\it number} of bound antibodies, one finds $F(x_s)=F(x_r)$ when the sample and reference numbers $x_s$ and $x_r$ are the same. Moreover, the strict monotonicity property $F$ implies the converse: 
\begin{align}
F(x_s)=F(x_r) \qquad {\rm implies} \qquad x_s = x_r. \label{eq:converse}
\end{align}
To relate these observations to concentration, we assume that $x$ is given by
\begin{align}
x=cd \label{eq:forward}
\end{align}
where $c$ is understood to be an undiluted concentration, and $d$ is a known dilution factor (with units of volume) typically having a numerical value  $d \le 1$ unit volume.\footnote{The parameter $d$ can of course be greater than unity, but this is interpreted as a sample being made more concentrated, which is not typical in experimental settings.}  
Taken together, Eqs.\ \eqref{eq:converse} and \eqref{eq:forward} suggests the following heuristic procedure for normalizing concentrations:
\begin{itemize}
\item[A.] Identify a reference material with a known (or assumed) concentration $c_{r,n}$.  Measure the fluorescence-dilution curve $F_r(d)$ by varying the dilution $d$.
\item[B.] For a sample $s$ with unknown concentration, measure the dilution curve $F_s(d)$. 
\item[C.] Find a value of $\alpha_{s}$ such that $F_r(d) = F_s(\alpha_{s} d)$ for all $d$.  
\item[D.] Set $\hat c_{s,n,r} = c_{r,n}/\alpha_{s}$.  
\end{itemize}
[{\it Nota bene:} In this section, we consider a fixed reference and assay.  Thus, for simplicity of notation, we sometimes suppress dependence of $\alpha_s$ on $n$ and $r$.  However, the $\alpha_s$ must be determined separately for each combination of the sample, reference, and assay.]

To rigorously justify this algorithm, note that the reference fluorescence $F_r(d)$  is determined by $F(x)$; i.e.\ $F_r(d)=F(c_{r,n}d)$ where $x=c_{r,n}d$ is the corresponding number of antibodies.  Likewise, $F_s(\alpha_s d) = F(c_{s,n} \alpha_s d)$.  Thus, by Eq.\ \eqref{eq:converse}, $F_r(d) = F_s(\alpha_s d) = F(x)$ implies the equivalent statements 
\begin{align}
\frac{c_{s,n}}{c_{r,n}} = \frac{1}{\alpha_{s,n,r}} = \hat y_{s,n,r} \iff \hat c_{s,n,r} \equiv c_{s,n} = \frac{c_{r,n}}{\alpha_{s,n,r}} \label{eq:equivalence}
\end{align}
The equality on the left-hand side (LHS) demonstrates that $\alpha_{s,n,r}^{-1}$ is the relative (bound) concentration of sample and reference antibodies when both are expressed in the same but arbitrary units.  The equality on the right-hand size (RHS) shows that when the units of $c_{r,n}$ are fixed, we may automatically interpret the antibody concentration $c_{s,n}$ as being normalized on the same scale.  Moreover, because Eqs.\ \eqref{eq:equivalence} are independent of $d$, the curves $F_r(d)$ and $F_s(\alpha_s d)$ should entirely overlap.  Thus, we can use this property to find $\hat c_{s,n,r}$, i.e.\ by solving for the value of $\alpha_s$ that collapses the $s$th dilution series onto $F_r(d)$.  See also Fig.\ \ref{fig:motivation} and Refs.\ \cite{qPCR,PCR2}.

In practice, estimating the $\alpha_s$ in this way is complicated by three issues that impact how we  formulate steps A through D.

First, measurements are often taken at a few serial dilutions $d_i$ ($i=1,2,...,D$) spanning several decades.  Thus $F(x)$ tends to be given on a sparse grid whose characteristic spacing grows exponentially.  To make the spacing more uniform, we take a logarithm of $F$ and express the resulting function in terms of ${\cal x}=\ln(x)$, which transforms the measurement domain to $-\infty < {\cal x} < \infty$.  This transformation also preserves strict monotonicity of $f({\cal x})=\ln(F(e^{\cal x})/U_F)$.  As an added benefit, we find that $f({\cal x})$ is typically sigmoidal.  That is, for some inflection point ${\cal x}_I$,  $f({\cal x})$ is convex (concave) when ${\cal x} \le {\cal x}_I$ (${\cal x}\ge {\cal x}_I$).  While not strictly necessary, this assumption is so convenient that we leverage it throughout.  Section \ref{sec:discussion} proposes generalizations and limitations of this choice.   

Second, the sparsity of the $d_i$ means that $f({\cal x})$ is only known at a few points, which makes it challenging to determine whether two dilution series coincide.  We address this problem by determining multiple $\alpha_s$ simultaneously by requiring that they all fall on the same curve.  Here the sigmoid structure of $f({\cal x})$ plays an important role by ensuring that this curve has a physically reasonable structure.

Third, we anticipate that the $\alpha_s$ are to be determined by some numerical method that iteratively varies these parameters to find their optimal values.  However, doing so makes the grid of ${\cal x}$ values dependent on the $\alpha_s$.  This motivates us to treat the fluorescence values of each  measurement as the independent variables, since these always define a fixed grid.  By the strict monotonicity of $f({\cal x})$, we may then write ${\cal x}={\cal x}(f)$ as a function of the fluorescence values, which effectively ``flips'' our perspective about the line ${\cal x} = f$.  Recalling that ${\cal x}=\ln(x) = \ln(cd)$, we now see that the equality 
\begin{align}
{\cal x} = \ln(\hat c_{s,n,r}d) &= \ln(c_{r,n}d/\alpha_s) \nonumber \\
&=\ln(c_{r,n}d) - \ln(\alpha_s)\label{eq:gameq}
\end{align} 
reinterprets $\ln(\alpha_s)$ as a constant vertical offset accounting for the difference between a reference and sample dilution series.

To realize these ideas mathematically, let $f_{i,r}$ be the fluorescence measurements associated with the reference material at dilution $d_i$.  Assume $S$ samples indexed by $s$ having unknown normalized concentrations $\hat c_{s,n,r}$.  For each of these samples, we assume corresponding measurements $f_{i,s}$ for dilutions $d_i$.  In practice, the dilutions can be different for each sample, although for simplicity we assume the same set for each sample.  Generalizations are trivial and left for the reader.

From this data, we create a single vector of elements $f_j$ comprised of the $f_{i,s}$ and $f_{i,r}$ in ascending order and without regard to their sample number or status as a reference. It is not problematic if values of $f_j$ are repeated.  Also let $\alpha_{s_j}$ and $d_{i_j}$ denote the corresponding value of $\alpha$ and $d$ for the $j$th fluorescence value, where $s_j$ can be a specific value of $s$ or denote the reference $r$.  To find the $\alpha_{s_j}$, we postulate the existence of true log-antibody numbers $\hat {\cal x}_j$, which should be sufficiently close to the values predicted by the $ c_{r,n}d_{i_j}/\alpha_{s_j}$.  In fact, under noiseless conditions, one expects
\begin{align}
\hat {\cal x}_j  - \ln(c_{r,n}d_{i_j}) + \ln(\alpha_{s_j}) = 0.
\end{align}
In practice, there will be noise, which suggests the objective
\begin{align}
\hat \L = \sum_j \left[ \hat {\cal x}_j - \ln(c_{r,n}d_{i_j}) + \ln(\alpha_{s_j}) \right]^2. \label{eq:objective}
\end{align}
Assuming a value for $c_{r,n}$ (which we can take to be $1$ for convenience), we minimize Eq.\ \eqref{eq:objective} as a function of the $\hat {\cal x}_j$ and the $\alpha_{s_j}$, subject to the constraints that
\begin{align}
\hat {\cal x}_j &= \hat {\cal x}_{j'} & {\rm if}  \qquad f_j &= f_{j'} \label{eq:firstequality} \\
\alpha_{s_j} &= 1 & {\rm if}  \qquad s_j &= r. \label{eq:secondequality}
\end{align}

\begin{figure}
\includegraphics[width=8cm]{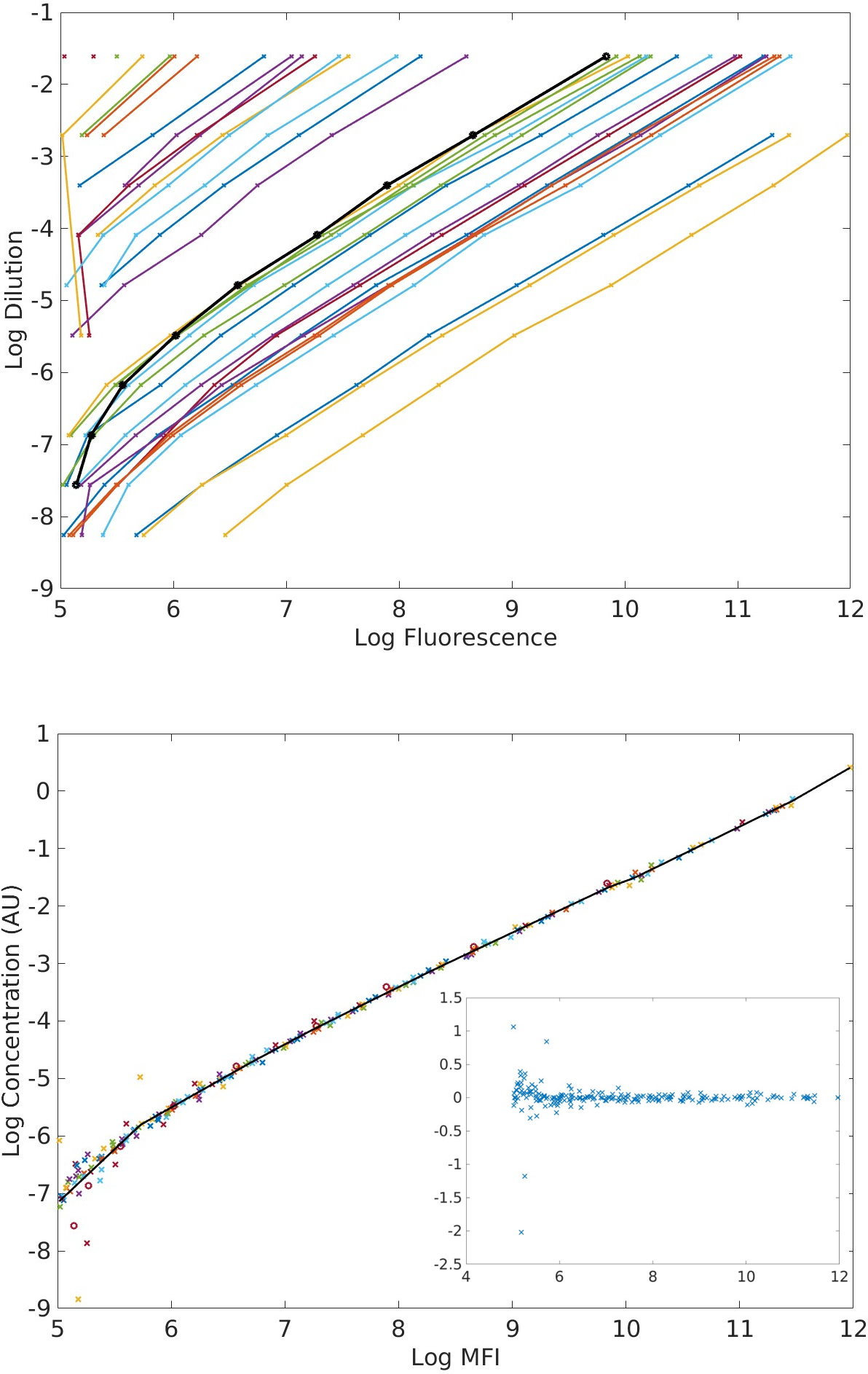}\caption{Raw data and data collapse associated with 34 SARS-CoV-2 positives samples and a mAb reference material, all measured via a ligand binding assay.  See Ref.\ \cite{InterlabLili} for experimental details.  {\it Top:}  Raw data associated with the samples.  The reference is labeled with the black circles.  Lines are guides for the eye, whereas the overlaid discrete data points are the dilution-fluorescence pairs measured by the instrument.  The axes are flipped relative to Fig.\ \ref{fig:motivation}.  By eye, it is plausible that a vertical shift applied to each dilution curve is sufficient to collapse them.  {\it Bottom:} Data collapse associated with minimizing Eq.\ \eqref{eq:regularized_objective}.  The reference material is assumed to have a dimensionless concentration of $1$.  The solid black curve is a reconstruction of the function $f({\cal x})$; see Eq. \eqref{eq:gameq} and surrounding text.  Note that the reference material has the average (weakly) non-linear behavior of the sample data after transformation.  The inset shows the residuals defined here as the difference between the transformed raw data and the estimated dilution curve in black.  }\label{fig:realdata}
\end{figure}

By itself, Eq.\ \eqref{eq:objective} does not define a well-posed optimization problem.  For example, if the $f_j$ all correspond to  distinct samples, then any values of the $\alpha_{s_j}$ will yield $\hat {\cal x}_j$ that yield $\L=0$.  {\it In such cases, we require that our analysis reduce to single-point normalizations based on Eq.\ \eqref{eq:old_norm}, which enforces theoretical consistency with past work, e.g.\ Refs.\ \cite{Endpoint,singlepoint}.}  We therefore add two constraints and a small regularization.  First, letting $j_k$ denote indices associated with unique values of $f_j$ in ascending order, we require that
\begin{align}
\hat {\cal x}_{j_{k+1}} \ge \hat {\cal x}_{j_k}. \label{eq:monotonicity}
\end{align}
That is, the antibody levels must be increasing with increasing fluorescence.  Second, we require that ${\cal x}(f)$ be concave for $f_{j_k}$ up to some inflection point $f_I$ and convex for all $f_{j_k} \ge f_I$.  This is equivalent to the sigmoid assumption transformed about the line ${\cal x}=f$.  To enforce this constraint, we construct a second-order, finite difference matrix $A_{m,j_k}(p)$ in terms of an undetermined inflection index $p$ using the procedure in the appendix of Ref.\ \cite{flowmeter}, where
\begin{subequations}
\begin{align}
\sum_{j_k}A_{m,j_k}{\cal x}_{j_k} &\le 0 & m \le p \label{eq:concave} \\
\sum_{j_k}A_{m,j_k}{\cal x}_{j_k} &\ge 0 & m > p. \label{eq:convex}
\end{align}
\end{subequations}
Third, we modify the objective function to be
\begin{align}
\hat {\mathcal L} \to \mathcal L = \hat {\mathcal L} &+ \tilde \epsilon_1 \sum_{k=k_{\rm low}}^{k_{\rm high}} \left(\frac{{\cal x}_{j_{k+1}}-{\cal x}_{j_{k-1}}}{f_{j_{k+1}}-f_{j_{k-1}}} -1 \right)^2 \nonumber \\
 &+ \tilde \epsilon_2 \sum_{m=1}^{N_{j_k}}\left(\sum_{j_k} A_{m,j_k}{\cal x}_{j_k} \right)^2 \label{eq:regularized_objective}
\end{align}
where $\tilde \epsilon_1$ and $\tilde \epsilon_2$ are small regularization parameters, and $k_{\rm low}$, $k_{\rm  high}$ are user-defined lower and upper limits between which we expect the fluorescence signals to be approximately linear with antibody number.  Thus, the regularization term associated with $\tilde \epsilon_1$ ensures that the reconstructed dilution curve has a linear region when there is only one measurement per sample.  The regularization associated with $\tilde \epsilon_2$ penalizes excessive curvature.  These parameters are chosen to have values that are roughly three decades smaller than the characteristic value of $\hat {\mathcal L}$ near its minimum, or, if $\hat {\mathcal L}=0$ is in the feasible set, we define $\tilde \epsilon_1=\tilde \epsilon_2=10^{-3}$.

To determine the remaining parameters, we minimize Eq.\ \eqref{eq:regularized_objective} with respect to the $\hat {\cal x}_j$, $\alpha_{s_j}$, and $p$, subject to Eqs.\ \eqref{eq:firstequality}--\eqref{eq:secondequality} and the inequality constraints \eqref{eq:monotonicity}--\eqref{eq:convex}.  It is straightforward to show that when the data is noiseless and the dilution curve is linear, the minimum of Eq.\ \eqref{eq:regularized_objective} is unique and yields the true values of $\alpha_s$ and ${\cal x}$.  Thus, our normalization procedure generalizes the techniques in Refs.\ \cite{Endpoint,singlepoint} and reduces to these approaches when only analyzing a single dilution associated with each sample and reference.

Figure \ref{fig:realdata} illustrates the results of this analysis applied to a collection of 38 SARS-CoV-2 positives samples and a mAb reference material, all measured using a ligand binding assay.  For this analysis we set $\tilde \epsilon_1=\tilde \epsilon_2 = 10^{-3}$ and $k_{\rm low}=k_{\rm high}$ to be the index associated with the measurement closest to the median fluorescence.  By eye, it is clear that the raw dilution curves all have the same approximate shape (top subplot).  After collapse, we find that with the exception of a few low-fluorescence data points, the characteristic deviation from the estimated dilution curve ${\cal x}$ is less than 5 \%, which is well within characteristic uncertainties associated with pipetting and sample preparation.  We speculate that the few data points showing significant deviation are exhibiting noise associated with being near the instrument noise floor.  Note also that we do not need to specify either a linear range or functional form of the dilution curve.  See Ref.\ \cite{InterlabLili} for more examples of this analysis applied to an interlab study with multiple distinct reference materials and assays.  

As a cautionary remark, the low-fluorescence data of Fig.\ \ref{fig:realdata} reveals the challenges of dealing with data fully at the noise-floor or upper saturation threshold.  In such cases, the amount of relevant physical information is dwarfed by instrument artifacts, which violates the assumptions underlying the optimization.  Thus, while our analysis does not need a linear fluorescence region {\it per se}, it does require the signal-to-noise ratio of the data to be suitably large.  To ensure this is the case, we remove from our analysis all data-points for which the signal-to-noise is less than roughly $1/10$.  While this choice is subjective, we find for the examples herein that it yields reasonable results.  

\section{Uncertainty Quantification}
\label{sec:betweenuc}

Minimization of Eq.\ \eqref{eq:regularized_objective} yields normalized antibody levels for a collection of samples and a reference material.  However, it does not by itself estimate the quantities $\Delta g_{r,n}$, $\Delta g_{s,n}$, or $\delta_{s,n,r}$.  We now address these tasks.

\begin{figure*}
\includegraphics[width=17cm]{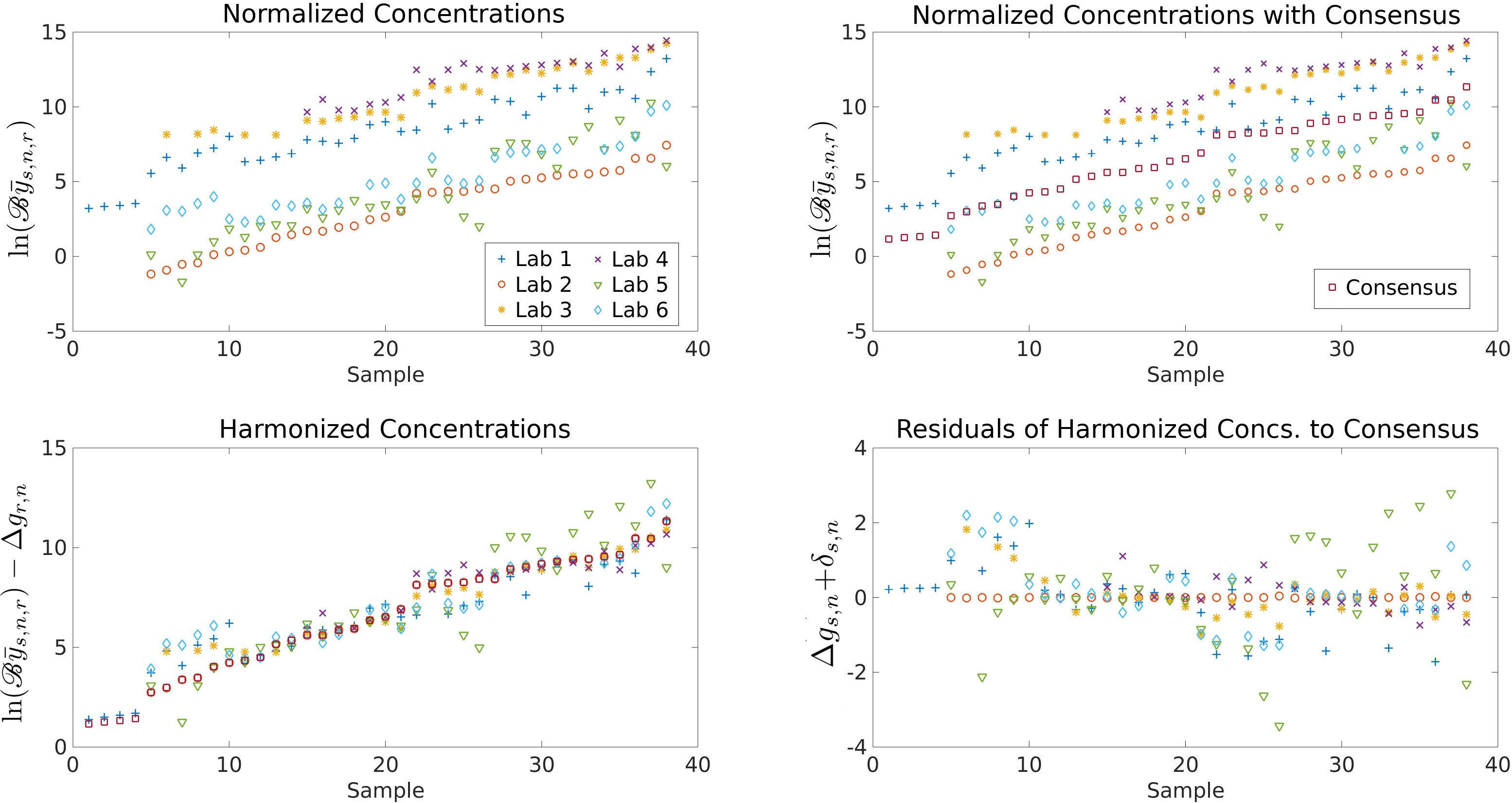}\caption{Example of the analysis leading to the solution given by Eq.\ \eqref{eq:optimal_solution}.  Symbols have the same meanings in all figures and are defined in the top plots.  {\it Top left:} Normalized data for 38 positive samples measured via 6 different assays.  {\it Top right:} Normalized data with consensus estimates given by Eq.\ \eqref{eq:optimal_solution}.  {\it Bottom left:} Harmonized data, i.e.\ normalized data corrected for the assay-dependent biases $\Delta g_{r,n}$ so that all samples are distributed about the consensus values.  {\it Bottom right:} Difference between the harmonized data and consensus values.}\label{fig:mab1}
\end{figure*}

While there are a variety of methods for estimating $\delta_{s,n,r}$ (e.g.\ \cite{Bootstrap1,Bootstrap2}), we consider the common situation wherein measurements are repeated.  Extensions and alternatives are discussed in Sec.\ \ref{sec:discussion}.  Assume therefore that we are given $S$ samples, $A$ assays, and one reference material.  Each sample and reference are measured $\tau$ times for each assay, where $t\in \{1,2,...,\tau\}$ indexes these time-points.  We use the analysis of Sec.\ \ref{sec:affine} to normalize all antibody measurements relative to the reference dilution curve measured at the same time.  Denote the corresponding antibody levels $\tilde y_{s,n,r,t}$.  Assuming that these measurements are independent in the $t$ index, we construct the arithmetic mean estimator
\begin{align}
\ln(\bar y_{s,n,r}) = \frac{1}{\tau}\sum_{t=1}^{\tau} \ln(\tilde y_{s,n,r,t}) \label{eq:logavg}
\end{align}
and sample standard uncertainty \cite{GUM}
\begin{align}
\sigma_{s,n,r}^2 = \frac{1}{\tau(\tau-1)}\sum_{t=1}^{\tau} \left[\ln(\bar y_{s,n,r}) - \ln(\tilde y_{s,n,r,t}) \right]^2,
\end{align}
where we approximate the variance ${\rm Var}[\delta_{s,n,r}]= \sigma_{s,n,r}^2$.  Note that the estimate for $\bar y_{s,n,r}$ corresponds to a geometric mean of antibody concentrations.  See Eqs.\ \eqref{eq:withinuc} and \eqref{eq:fullmodel}.\footnote{Recall that the logarithm of concentration is linear in the Gibbs free energy.  Thus Eq.\ \eqref{eq:logavg} can also be viewed as an estimate of the average $\Delta G$.}  Given a few replicates $\tau$ per sample, we make the additional minimal assumption that $\delta_{s,n,r}$ is a mean-zero normal random variable with variance $\sigma_{s,n,r}^2$.

Returning to Eq.\ \eqref{eq:fullmodel},
\begin{align}
\ln(\bar y_{s,n,r}) - \Delta g_{r,n} + \Delta g_{s,n} + \delta_{s,n,r} = \ln({\cal y}_{s,r}), \nonumber 
\end{align}
assume that $r$ is fixed and observe that the scaled and consensus values $\bar y_{s,n,r}$ and ${\cal y}_{s,r}$ depend on the reference material $r$.  
Provisionally assume that the assay-sample dependent effects $\Delta g_{s,n}$ are independent of  $r$, which we check after-the-fact by varying the reference.

\begin{figure}
\includegraphics[width=8cm]{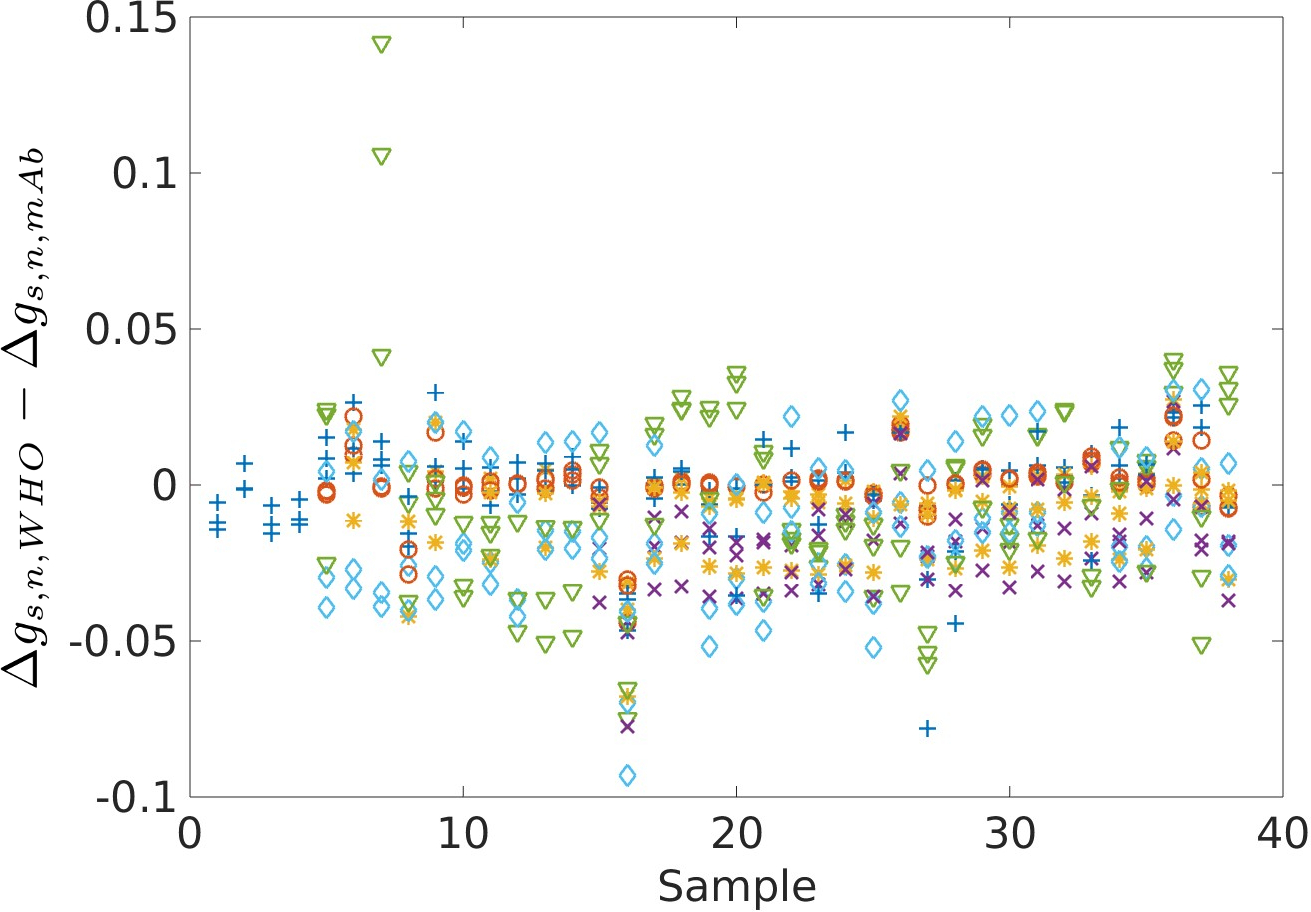}\caption{The difference of residuals $\Delta g_{s,n,WHO}-\Delta g_{s,n,mAb}$ computed from the WHO standard and the 3 other mAbs used in the interlab study associated with Ref.\ \cite{InterlabLili}.  Symbols and colors have the same meaning as in Fig.\ \ref{fig:mab1}.  Each point with a fixed color and symbol is associated with a different mAb.  The difference in residuals is typically less than 0.05, which corresponds to a relative variation in antibody number of 5 \%.  This demonstrates that the quantity $\Delta g_{s,n}$, which is associated with the sample-assay dependent randomness about the consensus value, does not depend on the reference material.}\label{fig:mabdiff}
\end{figure}

\begin{figure*}
\includegraphics[width=17cm]{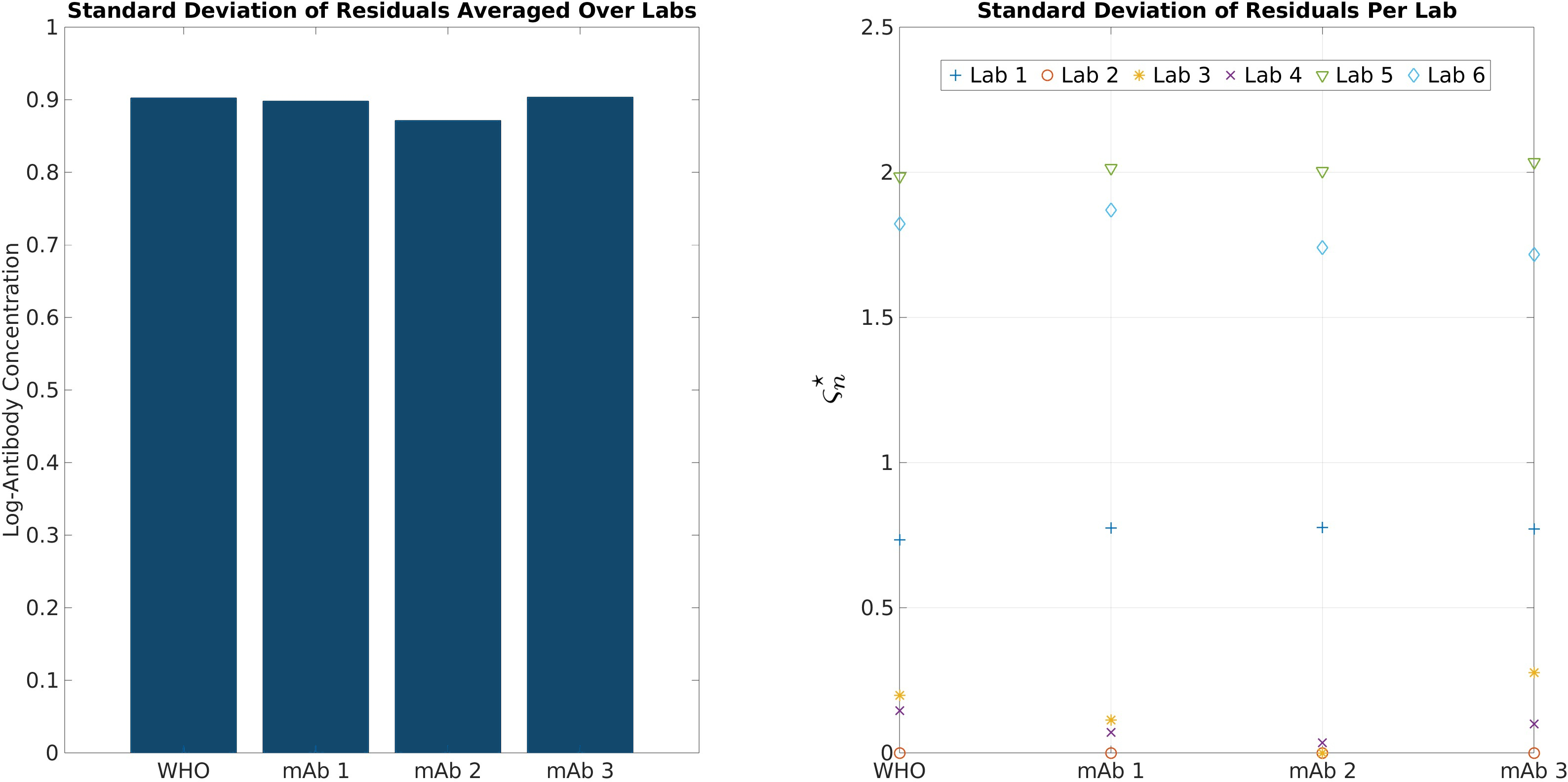}\caption{Further confirmation that approximate harmonization via Eq.\ \eqref{eq:optimal_solution} is reference independent.  {\it Left:} Square root of the average of $(\varsigma_n^\star)^2$ over all $n$ assays.  Note that the standard deviation is approximately constant across all standards.  {\it Right:} Maximum likelihood estimate of $\varsigma_n^\star$ as a function of assay.}\label{fig:varsigma}
\end{figure*}

Because the $\Delta g_{s,n}$ correspond to Gibbs Free-Energy changes associated with sample antibodies, we treat this quantity as an $s$-dependent realization of a mean-zero normal random variable with variance $\varsigma_n^2$.  That is, we assume
\begin{align}
\Delta g_{s,n}=\mathcal N_s\left(0,\varsigma_n^2\right). \label{eq:freedef}
\end{align}
Thus, the probability density of a value $\ln (\bar y_{s,n,r})$ is given by
\begin{align}
P \big (\ln(\bar y_{s,n,r})&|\Delta g_{r,n},{\cal y}_{s,r},\varsigma_n^2  \big ) \nonumber \\
& = \frac{e^{-\frac{\left(\ln({\cal y}_{s,r})-\Delta g_{r,n} - \ln(\bar y_{s,n,r})\right)^2}{2(\varsigma_n^2+\sigma_{s,n,r}^2)}}}{\sqrt{2\pi(\varsigma_n^2+\sigma_{s,n,r}^2)}}. \label{eq:indlike}
\end{align}
In the event that $\bar y_{s,n,r}$ falls below a detection threshold $y_\theta$ \textcolor{black}{(corresponding to dilution curves for which no measurement is above the signal-to-noise of $1/10$ in the examples above),} we can at most define the censored probability 
\begin{align}
\mathcal P(\bar y_{s,n,r} \le y_{\theta}) = \int_{-\infty}^{\ln(y_{\theta})}\hspace{-4mm}P \big (\bar z&|\Delta g_{r,n},{\cal y}_{s,r},\varsigma_n^2  \big ) {\rm d}\bar z \label{eq:censor}
\end{align}
in terms of Eq.\ \eqref{eq:indlike}, where we take $y_{\theta}$ to be the smallest measured value of $\ln(\bar y_{s,n,r}) > -\infty$; see also Refs.\ \cite{censor1,censor2}.  [For simplicity of notation, we have suppressed the dependence of $\mathcal P(\bar y_{s,n,r} \le y_{\theta})$ on $(\Delta g_{r,n},{\cal y}_{s,r},\varsigma_n^2)$.]  Given these quantities, we define a regularized, negative log-likelihood objective function to be
\begin{align}
\mathcal L_l &= - \sum_{\substack{s,n: \\ \bar y_{s,n,r} > y_{\theta}}}\ln \left[P\big (\ln(\bar y_{s,n,r})| \Delta g_{r,n},{\cal y}_{s,r},\varsigma_n^2  \big) \right] \nonumber \\ &\quad + \tilde \epsilon_3 \left[ \sum_{n} \Delta g_{r,n} \right]^2 \!\! - \!\! \sum_{\substack{s,n: \\ \bar y_{s,n,r} \le y_{\theta}}}\!\!\!\!\ln \left[\mathcal P(\bar y_{s,n,r} \le y_{\theta})\right] \label{eq:fulllike}
\end{align}
by summing Eq.\ \eqref{eq:indlike} over all samples and assays.  Observe that $\mathcal L_l$ is a function of the consensus values ${\cal y}_{s,r}$, reference-dependent free energy bias $\Delta g_{r,n}$, and assay-dependent free-energy variances $\varsigma_n^2$.  The value of $\tilde \epsilon_3$ can be set to any positive value (we set $\tilde \epsilon_3=1$), as the regularization only serves to remove a connected set associated with ambiguity in the value of $\Delta g_{r,n}$.  (That is, we can only determine the $\Delta g_{r,n}$ up to an additive constant, necessitating the regularization.)  Minimizing Eq.\ \eqref{eq:fulllike} yields estimates of these parameters via
\begin{align}
 \{\Delta g_{r,n}^\star\},\{{\cal y}_{s,r}^\star\},\{\varsigma_n^{2,\star}\}   =\underset{\{\Delta g_{r,n}\},\{{\cal y}_{s,r}\},\{\varsigma_n^{2}\}}{ {\rm argmin }} \!\!\! \mathcal L_l. \label{eq:optimal_solution}
\end{align}

While Eq.\ \eqref{eq:optimal_solution} is ultimately a choice of how to define consensus values, it admits an interpretation that is consistent with physical intuition.  In particular, recognize that $\Delta g_{s,n}$ (i.e.\ the sample-assay component of free energy) is treated as random because in general, we do not know {\it a priori} how a sample will interact with an assay.  In minimizing $\mathcal L_l$, we therefore seek the consensus values (up to a constant offset fixed by the regularization) and assay-dependent variability that maximizes the probability of the measured data.  Thus, we interpret Eq.\ \eqref{eq:optimal_solution} as the most likely probabilistic representation of the data \cite{RW}.  Moreover, including $\sigma_{s,n,r}^2$ in the definition of $\mathcal L_l$ accounts for the fact that measurements with high uncertainty contribute less to our knowledge of the corresponding consensus values and assay-dependent uncertainties.  See Sec.\ \ref{sec:discussion} for further analysis of this model.

Figure \ref{fig:mab1} shows the results of this analysis applied to mAb 1 considered in Ref.\ \cite{InterlabLili}.  The top plots show the normalized log-concentrations $\ln(\mathcal B \bar y_{s,n,r})$ with and without the consensus values for six different labs and 38 positive samples, where $\mathcal B = 1.2\times 10^6$ is the number of ng/mL of antibody in the undiluted reference.  The bottom plots show bias-corrected (i.e.\ harmonized) estimates  $\ln(\mathcal B \bar y_{s,n,r}) - \Delta g_{r,n}$.  The remaining variation is quantified by $\Delta g_{s,n} + \delta_{s,n}$ [recall Eq.\ \eqref{eq:separablenoise}].  Figure \ref{fig:mabdiff} shows the differences in residuals between harmonized concentrations and consensus values according to Eq.\ \eqref{eq:optimal_solution} using the WHO standard and three different mAbs described in Ref.\ \cite{InterlabLili}.  The bottom plot in particular shows that the residuals change by less than roughly 0.05 on a log scale when switching reference material, which corresponds to roughly 5\% relative variation in antibody concentration.  This validates that to within good approximation, the residuals can be expressed only in terms of $s$ and $n$ via $\Delta g_{s,n}+\delta_{s,n}$, as predicted by our thermodynamic model.  In other words, the lack of exact harmonization is due almost entirely to coordinated assay-sample effects.

To further illustrate this last point independently from the noise terms $\delta_{s,n}$ and $\delta_{s,n,r}$, recognize that the variance $(\varsigma_n^\star)^2$ is an estimate of the statistical properties of the $\Delta g_{s,n}$ via Eq.\ \eqref{eq:freedef}.  Figure \ref{fig:varsigma} shows the both the average of these variances over all $n$, as well as their assay-specific values.  All estimates are independent of the reference material used.

\section{Probabilistic Connection to Neutralization}
\label{sec:neuts}

The solution given by Eq.\ \eqref{eq:optimal_solution} can be used in Eq.\ \eqref{eq:indlike} to estimate the PDF $P({\cal y}_{s,r}|\bar y_{s,n,r},\Delta g_{r,n}^\star,(\varsigma_{n}^\star)^2)$ of a consensus value given the measured value.  This function can be used as the right-most term in the integrand of Eq.\ \eqref{eq:neutprob}.  In order to compute the integral, all that remains is to estimate the conditional probability density $P(\nu|{\cal y},r)$.  We make explicit the dependence on the reference $r$, since the consensus values depend on this parameter.

\begin{figure}
\includegraphics[width=8cm]{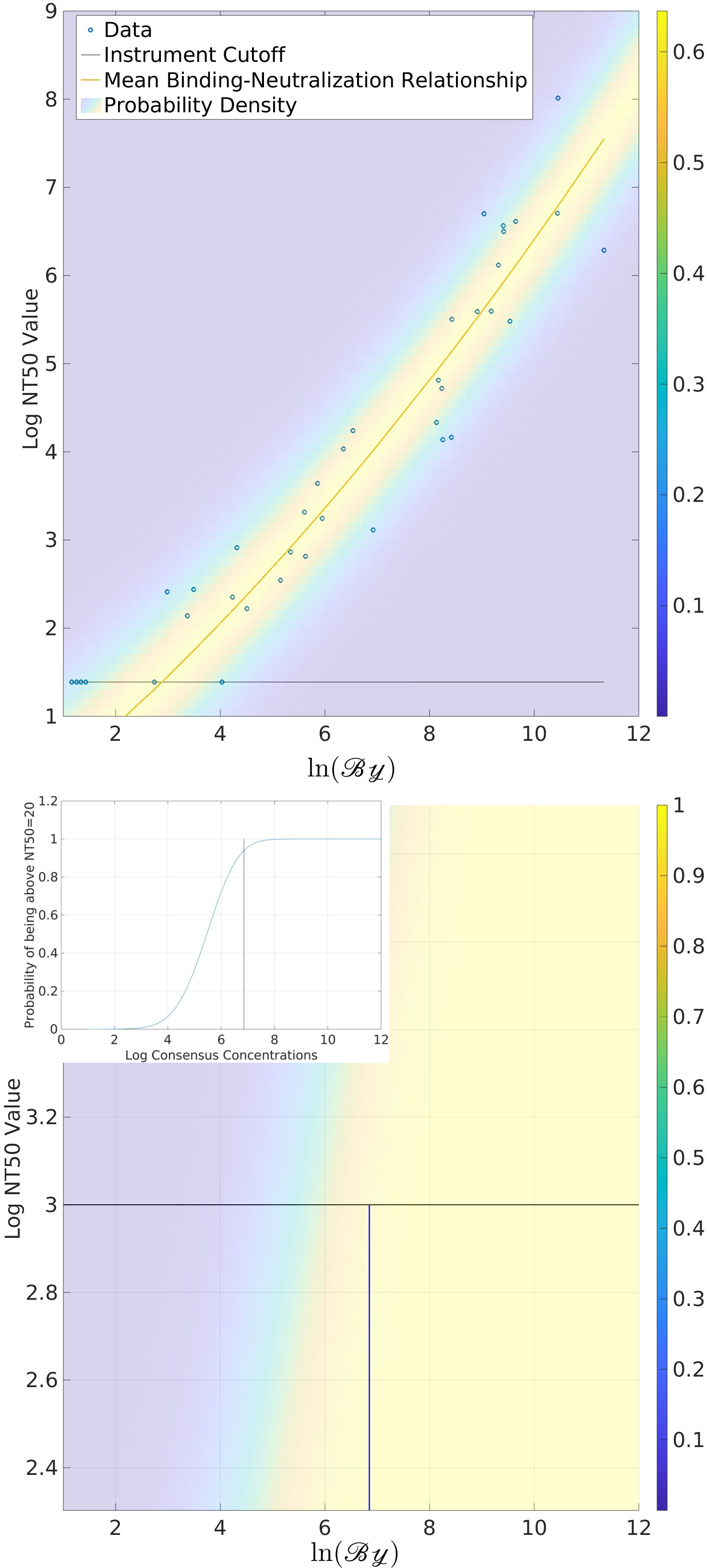}\caption{UQ for correlates of protection.  {\it Top:} Probability model $P(\nu|{\cal y},r)$ [expressed as $P(\ln(\nu)|\ln({\cal y}),r)$] linking log consensus binding levels relative to mAb 1 to log NT50 values for a psuedovirus assay developed at NIST; see Ref.\ \cite{InterlabLili}.  The horizontal black line is the cutoff below which NT50 were not reported, which we take to be the censoring level for the maximum likelihood estimation of model parameters.  The mean binding-neutralization relationship is assumed to be quadratic and have a noise structure given by Eq.\ \eqref{eq:neutmodel}.  {\it Bottom:} Integral of $P(\nu|{\cal y},r)$ with respect to $\nu$ for different log consensus binding levels (horizontal axis) and lower limits of integration (vertical axis).  The intersection of the black and blue lines is the binding level above which there is a greater than 95\% chance that the NT50 level is above 20; see Def.\ VI.  The inset shows the probability along the horizontal black line corresponding to this NT50 value.  }\label{fig:sherbert}
\end{figure}

Constructing $P(\nu|{\cal y},r)$ is an exercise in mathematical modeling that admits multiple paths forward.  For purposes of illustration, we consider a minimal approach wherein we assume that 
\begin{align}
\ln(\nu) &= \sum_{k=0}^K a_k \ln({\cal y})^k +\epsilon_{\nu,r} \label{eq:neutmodel} \\
\epsilon_{\nu,r} &= \mathcal N(0,\sigma_{\rm neut}^2)
\end{align}
is a polynomial of degree $K$ for some $K = \mathcal O(1)$ with additive, constant variance noise; see Fig.\ \ref{fig:sherbert} for data motivating this model.  For simplicity, we fit the $a_k$ and $\sigma_{\rm neut}^2$ to NT50, consensus sample pairs $(\nu_s,{\cal y}_{s,r})$ by minimizing the negative log-likelihood function, accounting for data censoring by analogy to Eq.\ \eqref{eq:censor} \cite{censor1,censor2}.  The bottom plot in this figure shows the integral
\begin{align}
I(\nu \ge \nu_{\min},{\cal y})= \int_{\nu_{\min}}^\infty P(\nu|{\cal y},r) {\rm d}\nu,
\end{align}
which is the probability of $\nu$ being greater than $\nu_{\min}$ given a consensus binding level. (In a slight abuse of notation, we express the integral in terms of $\nu$ and ${\cal y}$ but plot the probability as a function of logarithms thereof.)  This function permits us to identify, for example, a lower consensus binding threshold for which the neutralization level is above a minimum value with probability greater than 95\%; see Def.\ VI and the caption for more details.

To more fully connect the definition of a correlate of protection to a specific binding assay, it is necessary to directly compute the integral given by Eq.\ \eqref{eq:neutprob}.  In general, this task is assay specific.  However, given the probability model for $\Delta g_{s,n}$, it is straightforward to compute Eq.\ \eqref{eq:neutprob} via numerical means or in some cases, analytically in terms of polynomials of normal random variables.  Given the assay-specific nature of this task, we leave it for future work.

Despite the problem-specific nature of computing Eq.\ \eqref{eq:neutprob}, we can still deduce an important property of $P(\nu|\hat y_{s,n,r})$: it depends trivially on the reference via simple scale transformations.  To see this, note that the ratio $\hat y_{s,n,r}/\hat y_{s,n,r'} = c_{r',n}/c_{r,n}$ is sample independent.   Returning to Eq.\ \eqref{eq:neutprob}, one finds that
\begin{align}
P(\nu|\hat y_{s,n,r'}) = P(\nu | \hat y_{s,n,r}c_{r,n}/c_{r',n}).
\end{align}
While this result holds generally, in the context of Eq.\ \eqref{eq:neutmodel} it implies that the mean behavior of $\nu(\hat y_{s,n,r'})$ is rescaled in the $\hat y$ coordinate, or in the case of Fig.\ \ref{fig:sherbert}, translated laterally.  Thus, the choice of reference does not alter the information content (and thus usefulness) of $P(\nu | \hat y_{s,n,r})$, only its representation.  

\section{Discussion: Broader Implications for Serology}
\label{sec:discussion}

\subsection{Deeper Comparison with Past Works}

Several studies have considered the impacts of normalization and harmonization, both from the standpoint of establishing reference materials \cite{WHO1,WHO2,NCI} and deploying them in real-world settings \cite{Harmon1,Harmon2,Harmon3,Harmon4}.  However, none of these works formally defined the relationship between normalization and harmonization, implicitly taking these tasks to be identical.  More specifically, the authors tended to use manufacturer-specified scaling values (relative to a pooled standard) to harmonize measurements without the bias-correction term corresponding to $\Delta g_{r,n}$.  In terms of our analysis, this amounts to the assumption that $T(n,\hat c)=\hat c$; i.e.\ the equilibrium constants are separable.  

This lack of distinction between the concepts of harmonization and normalization may therefore be responsible for significant confusion within the serology community.  For example, a universal conclusion of Refs.\ \cite{Harmon1,Harmon2,Harmon3,Harmon4} has been that ``harmonized'' (i.e.\ normalized) measurements are not interchangeable, even when using human-derived standards.  Interestingly, the authors also observed at least two common trends: (i) normalized antibody measurements are correlated between different assays; and (ii) variability about a trend line increased with increasing antibody titers.  

Our analysis provides a likely explanation for both observed results.  Considering (i), the implicit choice $T(n,\hat c)=\hat c$ ignores the reference-dependent bias term $\exp(-\Delta g_{r,n})$ appearing in Eqs.\ \eqref{eq:bias_harmonization} and \eqref{eq:appox_harmonization_bias}.  Thus, it is not surprising that the harmonized values are proportional but not identical on average.  We predict this missing constant of proportionality should be $\exp(\Delta g_{r,n} - \Delta g_{r,n'})$ for two different assays $n$ and $n'$.  Concerning (ii), the increasing uncertainty with antibody concentrations is a direct manifestation of the uncertainty $\epsilon_{s,n}$ appearing in Eq.\ \eqref{eq:appox_harmonization_bias}, which we have already shown yields the {\it relative}  uncertainty $ T(n,\hat c_{s,n,r},r)\epsilon_{s,n}$ about consensus values.  We anticipate that a post-hoc analysis of the results in Refs.\ \cite{Harmon1,Harmon2,Harmon3,Harmon4} would lead to consistent predictions of the aforementioned physical quantities.  A potentially insurmountable, and thus disconcerting corollary is that the sample-assay dependent effects may be so large in real-world settings as to induce significant uncertainty, even using the harmonization techniques that we propose.

Another notable point of comparison is the calibration study performed in Ref.\ \cite{NCI}.  \textcolor{black}{The authors determined that in establishing the reference material, normalization via a human-derived, pooled standard harmonized samples to within a factor of two or better on average.}\footnote{In contrast, using different study designs and analyses, Refs.\ \cite{Harmon1,Harmon2,Harmon3,Harmon4} found that normalization still yielded up to 20-fold systematic discrepancies between assays .  Comparing these results to Ref.\ \cite{NCI} is challenging due to differences in the way results were reported.  Reference \cite{NCI} provided aggregate coefficients of variation across assays, whereas the other studies explicitly correlated normalized results between assays head-to-head.}  This contrasts with Ref.\ \cite{InterlabLili}, wherein we found that {\it after} correcting for reference-induced biases (i.e.\ $\Delta g_{r,n}$), harmonization could only be achieved to within a factor of approximately 2.5 ($\exp(0.9)$) on average; see Fig.\ \ref{fig:varsigma}.  Without the bias correction, harmonization was only achieved to within a factor of roughly 12 on average \cite{InterlabLili}.  The result of Ref.\ \cite{NCI} is especially surprising because the study therein included IgG measurements from SARS-CoV-2 spike, receptor binding domain (RBD), and nucleocapsid (N) assays, whereas we only considered the first two.  In the former study, one might expect that the corresponding set of $\Delta g_{r,n}$ and $\Delta g_{s,n}$ would be larger, considering that more types of assays were used.


A deeper analysis of the experimental design in Ref.\ \cite{NCI} suggests a resolution to this disparity.  It is noteworthy that the corresponding validation study only attempted to harmonize 5 samples (compared to our 38), four of which were also used to develop the standard.  In the context of Eq.\ \eqref{eq:regression_model}, it is likely that this choice of validation samples causes cancellation between the terms $\Delta g_{r,n}$ and $\Delta g_{s,n}$.  Ultimately, this would lead one to underestimate the true uncertainties associated with harmonization via normalization alone (i.e.\ without using our bias-correction).  Thus, we predict that a different validation study {\it not using samples also found in the standard} would yield results comparable to ours.  This prediction is consistent with Refs.\ \cite{Harmon1,Harmon2,Harmon3,Harmon4}.

\subsection{On the Mathematical Interpretation of a Consensus Value and Antibody Standard}
\label{subsec:interp}

It is notable that in Figs.\ \ref{fig:mab1} and \ref{fig:mabdiff}, the measurements from Lab 2 are nearly identical to the consensus values. This begs the obvious question of why the analysis yields such a result, and more generally, what is the interpretation of our consensus value?  

In this  instance, we note that the inter-day variation $\delta_{s,n,r}$ for Lab 2 is nearly a decade smaller than the corresponding uncertainties for the next closest lab.  Thus, it stands to reason that estimation of the consensus value is dominated by those measurements having the smallest uncertainties.  Ostensibly this is problematic: a precise measurement is not necessarily accurate.  However, we recall that antibody concentrations can only be estimated up to an unknown multiplicative factor, since the equilibrium constants are rarely, if ever determined; see Secs.\ \ref{subsec:units} and \ref{subsec:thermo}.

These observations suggest that a reasonable definition of consensus is one that minimizes disagreement between the results of different labs.  The maximum likelihood estimate given by Eq.\ \eqref{eq:optimal_solution} interprets disagreement as corresponding to low probability of the observed measurements under the assumption that the joint sample-assay component $\Delta g_{s,n}$ of the free energy is random.  [This modeling choice is justified by the fact that $\Delta g_{s,n}$ characterizes the immune response of an individual who is arbitrarily selected from a large population.]  In this context it is reasonable that a measurement with high uncertainty contributes less to the consensus estimate; its underlying true value may be closer to the remaining measurements than is reflected by its nominal value.


The subjectivity of this decision highlights the fact that from a purely performance standpoint, there may not exist a universal, best reference material for antibodies without further knowledge of equilibrium constants.  This stands in contrast to hardened metrological standards based on fundamental physical constants, e.g.\ as established by the connection between mass and Planck's constant \cite{Planck1,Planck2}.  Absent such relationships, we are forced to {\it choose} definitions for ``harmonization'' and ``consensus antibody value,'' and these necessarily control what we mean by a best standard.  In our analysis, the concept of ``best'' is associated with the reference that induces the least additional uncertainty (again a choice) into harmonization, which is in turn fixed by Def.\ V, Eq.\ \eqref{eq:optimal_solution}, and the supporting modeling assumptions.  However, our choices are not unique, and others may lead to distinct notions of a best standard.

A key challenge for the serology community is therefore to agree on a harmonization convention, which is necessary before standards can even be fully established.  The difficulty of addressing this problem is seen in Defs.\ IV and V.  They provide a generic structure of what harmonization entails, but critically, do not propose a functional form of the mapping $T(c,n)$ or noise $\epsilon_{s,n}$.  This latter task is complicated by virtue of being a task in mathematical modeling, although historically it has not be viewed as such.

\textcolor{black}{Here we propose that the modeling paradigm should be dictated by its usefulness.  In this respect, our approach has certain benefits grounded in its connection to physics.  Because the  underlying statistical model is induced by a thermodynamic description of antibody kinetics, it provides  an intuitive justification for various choices.  For example, Eq.\ \eqref{eq:logavg}, which is a geometric mean over antibody number, is revealed to be an arithmetic mean over Gibbs free-energies, a quantity for which this type of averaging may be appropriate.  Perhaps more importantly, the thermodynamics reveals how the reference-dependent bias can be removed as a source of uncertainty. Critically, the resulting equivalence of all standards for purposes of harmonization enables one to consider a broader definition of fitness of purpose.  Issues such as development times, manufacturing and distribution constraints, and traceability can become deciding factors in what constitutes a best serology standard.}

Ultimately the generality of Def.\ V permits multiple  interpretations of harmonization, and it is plausible that other approaches may further reduce uncertainty.  While we believe that our underlying approach is useful, our primary goal in this section and the previous is to highlight the {\it importance of rigorously defining and distinguishing the concepts of harmonization and normalization}.  These definitions and their realizations (e.g.\ via mathematical models) play a fundamental metrological role in ensuring reproducibility of measurements and developing reference materials.  

\subsection{On the Physical Interpretations of Gibbs Free-Energies and Consensus Values}

Antibodies in serology samples are typically polyclonal, as evidenced by the fact that distinct SARS-CoV-2 antigens (for example) can be detected in the same blood \cite{NCI,InterlabLili}.  Thus, the reaction process described by Eq.\ \eqref{eq:reaction} is a simplification of the true chemistry underlying serology assays.  A more accurate representation would consider a collection of simultaneous reactions for each type of antibody with associated reaction kinetics.  From a mathematical standpoint, however, this is problematic since one does not know {\it a priori} how many reactions to model.  Moreover, it is reasonable to assume that one type of antibody (or perhaps a small subset thereof) dominates the chemistry of a single assay, which justifies Eq.\ \eqref{eq:reaction}.

This suggests a need to re-interpret $\Delta g_{s,n}$.  In Sec.\ \ref{sec:perspectives}, this quantity represents the sample-assay specific contributions to the free-energy under the assumption that a single antibody interacts with the assay.  For a human-derived sample, we must add at a minimum that the specific type of antibody interacting with the assay can vary with the latter.  While this seems obvious -- e.g.\ some SARS-CoV-2 assays distinguish anti-nucleocapsid from anti-spike antibodies -- it weakens the concept of a consensus value.  That is, ${\chi}_{s,r}$ is not the total or even average concentration of antibodies of a specific type (e.g.\ anti-spike) in a sample.  At best, we can say that it is a characteristic concentration conditioned on the number and types of assays used to construct it.  In the context of Fig.\ \ref{fig:mab1}, for example, we might say that the consensus is the typical concentration of anti-SARS-CoV-2 antibodies across all types considered in Ref.\ \cite{InterlabLili}.

It is important to note that these observations do not change the underlying structure of our analysis, only its interpretation.  Likewise, these conclusions do not meaningfully change if we take the reference material to be a human-derived, pooled standard.  In such cases, we must re-interpret $\Delta g_{r,n}$ in the same way as we have done for $\Delta g_{s,n}$.  For human-derived standards, it is also reasonable that the raw, normalized antibody concentrations might exhibit less variation relative to the consensus as compared to mAbs, since the reference may have a collection of antibodies that will respond to each assay as might a test sample.  Indeed, this effect is evident in Ref.\ \cite{InterlabLili}.  {\it However}, as Figs.\ \ref{fig:mabdiff} and \ref{fig:varsigma} illustrate, this decrease in variance of the raw data does not impact the final uncertainty estimates of $\Delta g_{s,n}$ or reference-induced uncertainty, which we find to be constant.  Nor does it in any way change our harmonization method.

While the resulting interpretation of a consensus value suffers more ambiguity than hardened metrological concepts (e.g.\ fundamental physical constants), it is still {\it useful}.  This is illustrated by the analysis of Sec.\ \ref{sec:neuts}, which connects harmonized measurements to neutralization results via Def.\ VI.  UQ plays a critical role here: it allows us to quantify the extent to which a consensus value yields quantitative predictions about protection.  Our interpretation of these uncertainties as described above also points to methods for making consensus values more useful, e.g.\ by restricting the types of assays informing its defining so as to reduce ambiguity in what is meant by a consensus.

\subsection{Additional Limitations and Extensions}

The primary limitation of this work is the thermodynamic model inducing the probabilistic analysis.  Where the underlying physical assumptions are violated, our analysis may not be valid.  Examples are discussed in the previous section.  Other assays that may invalidate our analysis are those in which antibody avidity plays an important role, since the chemical reactions may be dominated by more complicated binding interactions not described by the Gibbs free-energy of Eq.\ \eqref{eq:Gibbs} \cite{avidity,Affinity2}.  However, the invariance of the residuals $\Delta g_{s,n} + \delta_{s,n}$ with respect to the reference provides a powerful tool that can be used to check the appropriateness of the assumptions, and thus our analysis.

Despite this limitation, our analyses provides several routes for generalization and/or incorporation of new physical information.  For example, the normalization procedure discussed in Sec.\ \ref{sec:affine} can be augmented with constraints to test for the degree of collapse among dilution curves or different assumptions about the structure of the underlying curve.  In the event that the residuals are too large, for example, one could hold out such samples for further investigation.  See Ref.\ \cite{qPCR} for related methods.  The probabilistic modeling of the within-lab uncertainties $\delta{s,n,r}$ can also be estimated via more sophisticated techniques, e.g.\ bootstrap-type methods \cite{Bootstrap1,Bootstrap2}

Finally, we observe that probabilistic modeling of the individual terms $\delta_{s,n}$ and $\delta_{r,n}$ appearing in Eq.\ \eqref{eq:separablenoise} may be useful in applications that seek to estimate correlates of protection in terms of $P({\cal y}_{s,r}|\bar y_{s,n,r},\Delta g_{r,n}^\star,(\varsigma_{n}^\star)^2)$.  While not needed to validate the analysis herein, the separability of $\delta_{s,n,r}$ implies that the $\Delta g_{r,n}$ have additional uncertainty associated the contribution from $\delta_{r,n}$.  Thus, it may be desirable to treat $\Delta g_{r,n}^\star$ as a random variable whose mean is given by our MLE analysis and distribution by the $\delta_{r,n}$.  Propagating the latter uncertainty into $P({\cal y}_{s,r}|\bar y_{s,n,r},\Delta g_{r,n}^\star,(\varsigma_{n}^\star)^2)$ would then provide more realistic estimates of this distribution.  

\subsection{Concluding Thoughts}

The main objective of this work is to provide a theoretical foundation for tasks such as antibody normalization, harmonization, and estimating correlates of protection.  As exercises in metrology, however, these tasks are challenging because their uncertainties are dominated by significant epistemic effects, e.g.\ lack of knowledge as to which antibodies are being detected, how they interact with the  measurement system, and even what we mean by an antibody concentration.  This is not a criticism so much as an observation: serology testing and immunity are difficult to understand due to complicated thermodynamic effects and inherent multiscale phenomena.  Our approach has therefore been to identify those aspects that cannot be made precise and leverage UQ as a means to quantify our lack of understanding.  This approach is enticing because it allows one to make informed decisions based on imperfect knowledge.  It also suggests routes for optimizing -- both informally and more mathematically -- aspects of serology testing, as well as diagnostics in general.  Looking forward then, our hope is that this work motivates a wider adoption of UQ within the biomedical community as a route to establishing rigorous principles of biometrology.

{\it Acknowledgements:} The authors thank Drs. Ronald Boisvert, Charles Romine, and especially Barry I.\ Schneider for helpful discussion during the preparation of this manuscript.    This manuscripts is an official contribution of the National Institute of Standards and Technology and is not subject to copyright in the United States.

{\it The NIST Research Protections Office has approved the use of data described herein.}

\Urlmuskip=0mu plus 1mu\relax

\bibliographystyle{apsrev4-1}
\bibliography{Interlab}

\begin{thebibliography}{55}%
\makeatletter
\providecommand \@ifxundefined [1]{%
 \@ifx{#1\undefined}
}%
\providecommand \@ifnum [1]{%
 \ifnum #1\expandafter \@firstoftwo
 \else \expandafter \@secondoftwo
 \fi
}%
\providecommand \@ifx [1]{%
 \ifx #1\expandafter \@firstoftwo
 \else \expandafter \@secondoftwo
 \fi
}%
\providecommand \natexlab [1]{#1}%
\providecommand \enquote  [1]{``#1''}%
\providecommand \bibnamefont  [1]{#1}%
\providecommand \bibfnamefont [1]{#1}%
\providecommand \citenamefont [1]{#1}%
\providecommand \href@noop [0]{\@secondoftwo}%
\providecommand \href [0]{\begingroup \@sanitize@url \@href}%
\providecommand \@href[1]{\@@startlink{#1}\@@href}%
\providecommand \@@href[1]{\endgroup#1\@@endlink}%
\providecommand \@sanitize@url [0]{\catcode `\\12\catcode `\$12\catcode
  `\&12\catcode `\#12\catcode `\^12\catcode `\_12\catcode `\%12\relax}%
\providecommand \@@startlink[1]{}%
\providecommand \@@endlink[0]{}%
\providecommand \url  [0]{\begingroup\@sanitize@url \@url }%
\providecommand \@url [1]{\endgroup\@href {#1}{\urlprefix }}%
\providecommand \urlprefix  [0]{URL }%
\providecommand \Eprint [0]{\href }%
\providecommand \doibase [0]{http://dx.doi.org/}%
\providecommand \selectlanguage [0]{\@gobble}%
\providecommand \bibinfo  [0]{\@secondoftwo}%
\providecommand \bibfield  [0]{\@secondoftwo}%
\providecommand \translation [1]{[#1]}%
\providecommand \BibitemOpen [0]{}%
\providecommand \bibitemStop [0]{}%
\providecommand \bibitemNoStop [0]{.\EOS\space}%
\providecommand \EOS [0]{\spacefactor3000\relax}%
\providecommand \BibitemShut  [1]{\csname bibitem#1\endcsname}%
\let\auto@bib@innerbib\@empty
\bibitem [{Ser()}]{Seronet}%
  \BibitemOpen
  \href@noop {} {\enquote {\bibinfo {title} {Serological sciences network},}\
  }\bibinfo {howpublished}
  {\url{https://www.cancer.gov/research/key-initiatives/covid-19/coronavirus-research-initiatives/serological-sciences-network}},\
  \bibinfo {note} {accessed: 2022-10-11}\BibitemShut {NoStop}%
\bibitem [{\citenamefont {Krammer}(2021)}]{Correlates1}%
  \BibitemOpen
  \bibfield  {author} {\bibinfo {author} {\bibfnamefont {F.}~\bibnamefont
  {Krammer}},\ }\href {\doibase 10.1016/S0140-6736(21)00782-0} {\bibfield
  {journal} {\bibinfo  {journal} {The Lancet}\ }\textbf {\bibinfo {volume}
  {397}},\ \bibinfo {pages} {1421} (\bibinfo {year} {2021})}\BibitemShut
  {NoStop}%
\bibitem [{\citenamefont {Feng}\ \emph {et~al.}(2021)\citenamefont {Feng},
  \citenamefont {Phillips}, \citenamefont {White}, \citenamefont {Sayal},
  \citenamefont {Aley}, \citenamefont {Bibi}, \citenamefont {Dold},
  \citenamefont {Fuskova}, \citenamefont {Gilbert}, \citenamefont {Hirsch},
  \citenamefont {Humphries}, \citenamefont {Jepson}, \citenamefont {Kelly},
  \citenamefont {Plested}, \citenamefont {Shoemaker}, \citenamefont {Thomas},
  \citenamefont {Vekemans}, \citenamefont {Villafana}, \citenamefont {Lambe},
  \citenamefont {Pollard}, \citenamefont {Voysey}, \citenamefont {Adlou},
  \citenamefont {Allen}, \citenamefont {Angus}, \citenamefont {Anslow},
  \citenamefont {Asselin}, \citenamefont {Baker}, \citenamefont {Baker},
  \citenamefont {Barlow}, \citenamefont {Beveridge}, \citenamefont {Bewley},
  \citenamefont {Brown}, \citenamefont {Brunt}, \citenamefont {Buttigieg},
  \citenamefont {Camara}, \citenamefont {Charlton}, \citenamefont {Chiplin},
  \citenamefont {Cicconi}, \citenamefont {Clutterbuck}, \citenamefont
  {Collins}, \citenamefont {Coombes}, \citenamefont {Clemens}, \citenamefont
  {Davison}, \citenamefont {Demissie}, \citenamefont {Dinesh}, \citenamefont
  {Douglas}, \citenamefont {Duncan}, \citenamefont {Emary}, \citenamefont
  {Ewer}, \citenamefont {Felle}, \citenamefont {Ferreira}, \citenamefont
  {Finn}, \citenamefont {Folegatti}, \citenamefont {Fothergill}, \citenamefont
  {Fraser}, \citenamefont {Garlant}, \citenamefont {Gatcombe}, \citenamefont
  {Godwin}, \citenamefont {Goodman}, \citenamefont {Green}, \citenamefont
  {Hallis}, \citenamefont {Hart}, \citenamefont {Heath}, \citenamefont {Hill},
  \citenamefont {Hill}, \citenamefont {Jenkin}, \citenamefont {Kasanyinga},
  \citenamefont {Kerridge}, \citenamefont {Knight}, \citenamefont {Leung},
  \citenamefont {Libri}, \citenamefont {Lillie}, \citenamefont {Marinou},
  \citenamefont {McGlashan}, \citenamefont {McGregor}, \citenamefont {McInroy},
  \citenamefont {Minassian}, \citenamefont {Mujadidi}, \citenamefont {Penn},
  \citenamefont {Petropoulos}, \citenamefont {Pollock}, \citenamefont {Proud},
  \citenamefont {Provstgaard-Morys}, \citenamefont {Rajapaska}, \citenamefont
  {Ramasamy}, \citenamefont {Sanders}, \citenamefont {Shaik}, \citenamefont
  {Singh}, \citenamefont {Smith}, \citenamefont {Snape}, \citenamefont {Song},
  \citenamefont {Shrestha}, \citenamefont {Sutherland}, \citenamefont
  {Thomson}, \citenamefont {Turner}, \citenamefont {Webb-Bridges},
  \citenamefont {Wrin}, \citenamefont {Williams},\ and\ \citenamefont {the
  Oxford COVID Vaccine Trial~Group}}]{Correlates2}%
  \BibitemOpen
  \bibfield  {author} {\bibinfo {author} {\bibfnamefont {S.}~\bibnamefont
  {Feng}}, \bibinfo {author} {\bibfnamefont {D.~J.}\ \bibnamefont {Phillips}},
  \bibinfo {author} {\bibfnamefont {T.}~\bibnamefont {White}}, \bibinfo
  {author} {\bibfnamefont {H.}~\bibnamefont {Sayal}}, \bibinfo {author}
  {\bibfnamefont {P.~K.}\ \bibnamefont {Aley}}, \bibinfo {author}
  {\bibfnamefont {S.}~\bibnamefont {Bibi}}, \bibinfo {author} {\bibfnamefont
  {C.}~\bibnamefont {Dold}}, \bibinfo {author} {\bibfnamefont {M.}~\bibnamefont
  {Fuskova}}, \bibinfo {author} {\bibfnamefont {S.~C.}\ \bibnamefont
  {Gilbert}}, \bibinfo {author} {\bibfnamefont {I.}~\bibnamefont {Hirsch}},
  \bibinfo {author} {\bibfnamefont {H.~E.}\ \bibnamefont {Humphries}}, \bibinfo
  {author} {\bibfnamefont {B.}~\bibnamefont {Jepson}}, \bibinfo {author}
  {\bibfnamefont {E.~J.}\ \bibnamefont {Kelly}}, \bibinfo {author}
  {\bibfnamefont {E.}~\bibnamefont {Plested}}, \bibinfo {author} {\bibfnamefont
  {K.}~\bibnamefont {Shoemaker}}, \bibinfo {author} {\bibfnamefont {K.~M.}\
  \bibnamefont {Thomas}}, \bibinfo {author} {\bibfnamefont {J.}~\bibnamefont
  {Vekemans}}, \bibinfo {author} {\bibfnamefont {T.~L.}\ \bibnamefont
  {Villafana}}, \bibinfo {author} {\bibfnamefont {T.}~\bibnamefont {Lambe}},
  \bibinfo {author} {\bibfnamefont {A.~J.}\ \bibnamefont {Pollard}}, \bibinfo
  {author} {\bibfnamefont {M.}~\bibnamefont {Voysey}}, \bibinfo {author}
  {\bibfnamefont {S.}~\bibnamefont {Adlou}}, \bibinfo {author} {\bibfnamefont
  {L.}~\bibnamefont {Allen}}, \bibinfo {author} {\bibfnamefont
  {B.}~\bibnamefont {Angus}}, \bibinfo {author} {\bibfnamefont
  {R.}~\bibnamefont {Anslow}}, \bibinfo {author} {\bibfnamefont {M.-C.}\
  \bibnamefont {Asselin}}, \bibinfo {author} {\bibfnamefont {N.}~\bibnamefont
  {Baker}}, \bibinfo {author} {\bibfnamefont {P.}~\bibnamefont {Baker}},
  \bibinfo {author} {\bibfnamefont {T.}~\bibnamefont {Barlow}}, \bibinfo
  {author} {\bibfnamefont {A.}~\bibnamefont {Beveridge}}, \bibinfo {author}
  {\bibfnamefont {K.~R.}\ \bibnamefont {Bewley}}, \bibinfo {author}
  {\bibfnamefont {P.}~\bibnamefont {Brown}}, \bibinfo {author} {\bibfnamefont
  {E.}~\bibnamefont {Brunt}}, \bibinfo {author} {\bibfnamefont {K.~R.}\
  \bibnamefont {Buttigieg}}, \bibinfo {author} {\bibfnamefont {S.}~\bibnamefont
  {Camara}}, \bibinfo {author} {\bibfnamefont {S.}~\bibnamefont {Charlton}},
  \bibinfo {author} {\bibfnamefont {E.}~\bibnamefont {Chiplin}}, \bibinfo
  {author} {\bibfnamefont {P.}~\bibnamefont {Cicconi}}, \bibinfo {author}
  {\bibfnamefont {E.~A.}\ \bibnamefont {Clutterbuck}}, \bibinfo {author}
  {\bibfnamefont {A.~M.}\ \bibnamefont {Collins}}, \bibinfo {author}
  {\bibfnamefont {N.~S.}\ \bibnamefont {Coombes}}, \bibinfo {author}
  {\bibfnamefont {S.~A.~C.}\ \bibnamefont {Clemens}}, \bibinfo {author}
  {\bibfnamefont {M.}~\bibnamefont {Davison}}, \bibinfo {author} {\bibfnamefont
  {T.}~\bibnamefont {Demissie}}, \bibinfo {author} {\bibfnamefont
  {T.}~\bibnamefont {Dinesh}}, \bibinfo {author} {\bibfnamefont {A.~D.}\
  \bibnamefont {Douglas}}, \bibinfo {author} {\bibfnamefont {C.~J.~A.}\
  \bibnamefont {Duncan}}, \bibinfo {author} {\bibfnamefont {K.~R.~W.}\
  \bibnamefont {Emary}}, \bibinfo {author} {\bibfnamefont {K.~J.}\ \bibnamefont
  {Ewer}}, \bibinfo {author} {\bibfnamefont {S.}~\bibnamefont {Felle}},
  \bibinfo {author} {\bibfnamefont {D.~M.}\ \bibnamefont {Ferreira}}, \bibinfo
  {author} {\bibfnamefont {A.}~\bibnamefont {Finn}}, \bibinfo {author}
  {\bibfnamefont {P.~M.}\ \bibnamefont {Folegatti}}, \bibinfo {author}
  {\bibfnamefont {R.}~\bibnamefont {Fothergill}}, \bibinfo {author}
  {\bibfnamefont {S.}~\bibnamefont {Fraser}}, \bibinfo {author} {\bibfnamefont
  {H.}~\bibnamefont {Garlant}}, \bibinfo {author} {\bibfnamefont
  {L.}~\bibnamefont {Gatcombe}}, \bibinfo {author} {\bibfnamefont {K.~J.}\
  \bibnamefont {Godwin}}, \bibinfo {author} {\bibfnamefont {A.~L.}\
  \bibnamefont {Goodman}}, \bibinfo {author} {\bibfnamefont {C.~A.}\
  \bibnamefont {Green}}, \bibinfo {author} {\bibfnamefont {B.}~\bibnamefont
  {Hallis}}, \bibinfo {author} {\bibfnamefont {T.~C.}\ \bibnamefont {Hart}},
  \bibinfo {author} {\bibfnamefont {P.~T.}\ \bibnamefont {Heath}}, \bibinfo
  {author} {\bibfnamefont {H.}~\bibnamefont {Hill}}, \bibinfo {author}
  {\bibfnamefont {A.~V.~S.}\ \bibnamefont {Hill}}, \bibinfo {author}
  {\bibfnamefont {D.}~\bibnamefont {Jenkin}}, \bibinfo {author} {\bibfnamefont
  {M.}~\bibnamefont {Kasanyinga}}, \bibinfo {author} {\bibfnamefont
  {S.}~\bibnamefont {Kerridge}}, \bibinfo {author} {\bibfnamefont
  {C.}~\bibnamefont {Knight}}, \bibinfo {author} {\bibfnamefont
  {S.}~\bibnamefont {Leung}}, \bibinfo {author} {\bibfnamefont
  {V.}~\bibnamefont {Libri}}, \bibinfo {author} {\bibfnamefont {P.~J.}\
  \bibnamefont {Lillie}}, \bibinfo {author} {\bibfnamefont {S.}~\bibnamefont
  {Marinou}}, \bibinfo {author} {\bibfnamefont {J.}~\bibnamefont {McGlashan}},
  \bibinfo {author} {\bibfnamefont {A.~C.}\ \bibnamefont {McGregor}}, \bibinfo
  {author} {\bibfnamefont {L.}~\bibnamefont {McInroy}}, \bibinfo {author}
  {\bibfnamefont {A.~M.}\ \bibnamefont {Minassian}}, \bibinfo {author}
  {\bibfnamefont {Y.~F.}\ \bibnamefont {Mujadidi}}, \bibinfo {author}
  {\bibfnamefont {E.~J.}\ \bibnamefont {Penn}}, \bibinfo {author}
  {\bibfnamefont {C.~J.}\ \bibnamefont {Petropoulos}}, \bibinfo {author}
  {\bibfnamefont {K.~M.}\ \bibnamefont {Pollock}}, \bibinfo {author}
  {\bibfnamefont {P.~C.}\ \bibnamefont {Proud}}, \bibinfo {author}
  {\bibfnamefont {S.}~\bibnamefont {Provstgaard-Morys}}, \bibinfo {author}
  {\bibfnamefont {D.}~\bibnamefont {Rajapaska}}, \bibinfo {author}
  {\bibfnamefont {M.~N.}\ \bibnamefont {Ramasamy}}, \bibinfo {author}
  {\bibfnamefont {K.}~\bibnamefont {Sanders}}, \bibinfo {author} {\bibfnamefont
  {I.}~\bibnamefont {Shaik}}, \bibinfo {author} {\bibfnamefont
  {N.}~\bibnamefont {Singh}}, \bibinfo {author} {\bibfnamefont
  {A.}~\bibnamefont {Smith}}, \bibinfo {author} {\bibfnamefont {M.~D.}\
  \bibnamefont {Snape}}, \bibinfo {author} {\bibfnamefont {R.}~\bibnamefont
  {Song}}, \bibinfo {author} {\bibfnamefont {S.}~\bibnamefont {Shrestha}},
  \bibinfo {author} {\bibfnamefont {R.~K.}\ \bibnamefont {Sutherland}},
  \bibinfo {author} {\bibfnamefont {E.~C.}\ \bibnamefont {Thomson}}, \bibinfo
  {author} {\bibfnamefont {D.~P.~J.}\ \bibnamefont {Turner}}, \bibinfo {author}
  {\bibfnamefont {A.}~\bibnamefont {Webb-Bridges}}, \bibinfo {author}
  {\bibfnamefont {T.}~\bibnamefont {Wrin}}, \bibinfo {author} {\bibfnamefont
  {C.~J.}\ \bibnamefont {Williams}}, \ and\ \bibinfo {author} {\bibnamefont
  {the Oxford COVID Vaccine Trial~Group}},\ }\href {\doibase
  10.1038/s41591-021-01540-1} {\bibfield  {journal} {\bibinfo  {journal}
  {Nature Medicine}\ }\textbf {\bibinfo {volume} {27}},\ \bibinfo {pages}
  {2032} (\bibinfo {year} {2021})}\BibitemShut {NoStop}%
\bibitem [{\citenamefont {Perry}\ \emph {et~al.}(2022)\citenamefont {Perry},
  \citenamefont {Osman}, \citenamefont {Wright}, \citenamefont
  {Richard-Greenblatt}, \citenamefont {Buchan}, \citenamefont {Sadarangani},\
  and\ \citenamefont {Bolotin}}]{Correlates3}%
  \BibitemOpen
  \bibfield  {author} {\bibinfo {author} {\bibfnamefont {J.}~\bibnamefont
  {Perry}}, \bibinfo {author} {\bibfnamefont {S.}~\bibnamefont {Osman}},
  \bibinfo {author} {\bibfnamefont {J.}~\bibnamefont {Wright}}, \bibinfo
  {author} {\bibfnamefont {M.}~\bibnamefont {Richard-Greenblatt}}, \bibinfo
  {author} {\bibfnamefont {S.~A.}\ \bibnamefont {Buchan}}, \bibinfo {author}
  {\bibfnamefont {M.}~\bibnamefont {Sadarangani}}, \ and\ \bibinfo {author}
  {\bibfnamefont {S.}~\bibnamefont {Bolotin}},\ }\href {\doibase
  10.1371/journal.pone.0266852} {\bibfield  {journal} {\bibinfo  {journal}
  {PLOS ONE}\ }\textbf {\bibinfo {volume} {17}},\ \bibinfo {pages} {1}
  (\bibinfo {year} {2022})}\BibitemShut {NoStop}%
\bibitem [{\citenamefont {Abbasi}(2022)}]{Correlates4}%
  \BibitemOpen
  \bibfield  {author} {\bibinfo {author} {\bibfnamefont {J.}~\bibnamefont
  {Abbasi}},\ }\href {\doibase 10.1001/jama.2021.24117} {\bibfield  {journal}
  {\bibinfo  {journal} {JAMA}\ }\textbf {\bibinfo {volume} {327}},\ \bibinfo
  {pages} {115} (\bibinfo {year} {2022})}\BibitemShut {NoStop}%
\bibitem [{\citenamefont {Karger}\ \emph {et~al.}(2022)\citenamefont {Karger},
  \citenamefont {Brien}, \citenamefont {Christen}, \citenamefont {Dhakal},
  \citenamefont {Kemp}, \citenamefont {Klein}, \citenamefont {Pinto},
  \citenamefont {Premkumar}, \citenamefont {Roback}, \citenamefont {Binder},
  \citenamefont {Boehme}, \citenamefont {Boppana}, \citenamefont
  {Cordon-Cardo}, \citenamefont {Crawford}, \citenamefont {Daiss},
  \citenamefont {Dupuis}, \citenamefont {Espino}, \citenamefont
  {Firpo-Betancourt}, \citenamefont {Forconi}, \citenamefont {Forrest},
  \citenamefont {Girardin}, \citenamefont {Granger}, \citenamefont {Granger},
  \citenamefont {Haddad}, \citenamefont {Heaney}, \citenamefont {Hunt},
  \citenamefont {Kennedy}, \citenamefont {King}, \citenamefont {Krammer},
  \citenamefont {Kruczynski}, \citenamefont {LaBaer}, \citenamefont {Lee},
  \citenamefont {Lee}, \citenamefont {Liu}, \citenamefont {Lozanski},
  \citenamefont {Lucas}, \citenamefont {Mendu}, \citenamefont {Moormann},
  \citenamefont {Murugan}, \citenamefont {Okoye}, \citenamefont {Pantoja},
  \citenamefont {Payne}, \citenamefont {Park}, \citenamefont {Pinninti},
  \citenamefont {Pinto}, \citenamefont {Pisanic}, \citenamefont {Qiu},
  \citenamefont {Sariol}, \citenamefont {Simon}, \citenamefont {Song},
  \citenamefont {Steffen}, \citenamefont {Stone}, \citenamefont {Styer},
  \citenamefont {Suthar}, \citenamefont {Thomas}, \citenamefont {Thyagarajan},
  \citenamefont {Wajnberg}, \citenamefont {Yates},\ and\ \citenamefont
  {Sobhani}}]{Seronet2}%
  \BibitemOpen
  \bibfield  {author} {\bibinfo {author} {\bibfnamefont {A.~B.}\ \bibnamefont
  {Karger}}, \bibinfo {author} {\bibfnamefont {J.~D.}\ \bibnamefont {Brien}},
  \bibinfo {author} {\bibfnamefont {J.~M.}\ \bibnamefont {Christen}}, \bibinfo
  {author} {\bibfnamefont {S.}~\bibnamefont {Dhakal}}, \bibinfo {author}
  {\bibfnamefont {T.~J.}\ \bibnamefont {Kemp}}, \bibinfo {author}
  {\bibfnamefont {S.~L.}\ \bibnamefont {Klein}}, \bibinfo {author}
  {\bibfnamefont {L.~A.}\ \bibnamefont {Pinto}}, \bibinfo {author}
  {\bibfnamefont {L.}~\bibnamefont {Premkumar}}, \bibinfo {author}
  {\bibfnamefont {J.~D.}\ \bibnamefont {Roback}}, \bibinfo {author}
  {\bibfnamefont {R.~A.}\ \bibnamefont {Binder}}, \bibinfo {author}
  {\bibfnamefont {K.~W.}\ \bibnamefont {Boehme}}, \bibinfo {author}
  {\bibfnamefont {S.}~\bibnamefont {Boppana}}, \bibinfo {author} {\bibfnamefont
  {C.}~\bibnamefont {Cordon-Cardo}}, \bibinfo {author} {\bibfnamefont {J.~M.}\
  \bibnamefont {Crawford}}, \bibinfo {author} {\bibfnamefont {J.~L.}\
  \bibnamefont {Daiss}}, \bibinfo {author} {\bibfnamefont {A.~P.}\ \bibnamefont
  {Dupuis}}, \bibinfo {author} {\bibfnamefont {A.~M.}\ \bibnamefont {Espino}},
  \bibinfo {author} {\bibfnamefont {A.}~\bibnamefont {Firpo-Betancourt}},
  \bibinfo {author} {\bibfnamefont {C.}~\bibnamefont {Forconi}}, \bibinfo
  {author} {\bibfnamefont {J.~C.}\ \bibnamefont {Forrest}}, \bibinfo {author}
  {\bibfnamefont {R.~C.}\ \bibnamefont {Girardin}}, \bibinfo {author}
  {\bibfnamefont {D.~A.}\ \bibnamefont {Granger}}, \bibinfo {author}
  {\bibfnamefont {S.~W.}\ \bibnamefont {Granger}}, \bibinfo {author}
  {\bibfnamefont {N.~S.}\ \bibnamefont {Haddad}}, \bibinfo {author}
  {\bibfnamefont {C.~D.}\ \bibnamefont {Heaney}}, \bibinfo {author}
  {\bibfnamefont {D.~T.}\ \bibnamefont {Hunt}}, \bibinfo {author}
  {\bibfnamefont {J.~L.}\ \bibnamefont {Kennedy}}, \bibinfo {author}
  {\bibfnamefont {C.~L.}\ \bibnamefont {King}}, \bibinfo {author}
  {\bibfnamefont {F.}~\bibnamefont {Krammer}}, \bibinfo {author} {\bibfnamefont
  {K.}~\bibnamefont {Kruczynski}}, \bibinfo {author} {\bibfnamefont
  {J.}~\bibnamefont {LaBaer}}, \bibinfo {author} {\bibfnamefont {F.~E.-H.}\
  \bibnamefont {Lee}}, \bibinfo {author} {\bibfnamefont {W.~T.}\ \bibnamefont
  {Lee}}, \bibinfo {author} {\bibfnamefont {S.-L.}\ \bibnamefont {Liu}},
  \bibinfo {author} {\bibfnamefont {G.}~\bibnamefont {Lozanski}}, \bibinfo
  {author} {\bibfnamefont {T.}~\bibnamefont {Lucas}}, \bibinfo {author}
  {\bibfnamefont {D.~R.}\ \bibnamefont {Mendu}}, \bibinfo {author}
  {\bibfnamefont {A.~M.}\ \bibnamefont {Moormann}}, \bibinfo {author}
  {\bibfnamefont {V.}~\bibnamefont {Murugan}}, \bibinfo {author} {\bibfnamefont
  {N.~C.}\ \bibnamefont {Okoye}}, \bibinfo {author} {\bibfnamefont
  {P.}~\bibnamefont {Pantoja}}, \bibinfo {author} {\bibfnamefont {A.~F.}\
  \bibnamefont {Payne}}, \bibinfo {author} {\bibfnamefont {J.}~\bibnamefont
  {Park}}, \bibinfo {author} {\bibfnamefont {S.}~\bibnamefont {Pinninti}},
  \bibinfo {author} {\bibfnamefont {A.~K.}\ \bibnamefont {Pinto}}, \bibinfo
  {author} {\bibfnamefont {N.}~\bibnamefont {Pisanic}}, \bibinfo {author}
  {\bibfnamefont {J.}~\bibnamefont {Qiu}}, \bibinfo {author} {\bibfnamefont
  {C.~A.}\ \bibnamefont {Sariol}}, \bibinfo {author} {\bibfnamefont
  {V.}~\bibnamefont {Simon}}, \bibinfo {author} {\bibfnamefont
  {L.}~\bibnamefont {Song}}, \bibinfo {author} {\bibfnamefont {T.~L.}\
  \bibnamefont {Steffen}}, \bibinfo {author} {\bibfnamefont {E.~T.}\
  \bibnamefont {Stone}}, \bibinfo {author} {\bibfnamefont {L.~M.}\ \bibnamefont
  {Styer}}, \bibinfo {author} {\bibfnamefont {M.~S.}\ \bibnamefont {Suthar}},
  \bibinfo {author} {\bibfnamefont {S.~N.}\ \bibnamefont {Thomas}}, \bibinfo
  {author} {\bibfnamefont {B.}~\bibnamefont {Thyagarajan}}, \bibinfo {author}
  {\bibfnamefont {A.}~\bibnamefont {Wajnberg}}, \bibinfo {author}
  {\bibfnamefont {J.~L.}\ \bibnamefont {Yates}}, \ and\ \bibinfo {author}
  {\bibfnamefont {K.}~\bibnamefont {Sobhani}},\ }\href {\doibase
  10.1101/2022.02.27.22271399} {\bibfield  {journal} {\bibinfo  {journal}
  {medRxiv}\ } (\bibinfo {year} {2022}),\
  10.1101/2022.02.27.22271399}\BibitemShut {NoStop}%
\bibitem [{\citenamefont {Kristiansen}\ \emph {et~al.}(2021)\citenamefont
  {Kristiansen}, \citenamefont {Page}, \citenamefont {Bernasconi},
  \citenamefont {Mattiuzzo}, \citenamefont {Dull}, \citenamefont {Makar},
  \citenamefont {Plotkin},\ and\ \citenamefont {Knezevic}}]{WHO1}%
  \BibitemOpen
  \bibfield  {author} {\bibinfo {author} {\bibfnamefont {P.~A.}\ \bibnamefont
  {Kristiansen}}, \bibinfo {author} {\bibfnamefont {M.}~\bibnamefont {Page}},
  \bibinfo {author} {\bibfnamefont {V.}~\bibnamefont {Bernasconi}}, \bibinfo
  {author} {\bibfnamefont {G.}~\bibnamefont {Mattiuzzo}}, \bibinfo {author}
  {\bibfnamefont {P.}~\bibnamefont {Dull}}, \bibinfo {author} {\bibfnamefont
  {K.}~\bibnamefont {Makar}}, \bibinfo {author} {\bibfnamefont
  {S.}~\bibnamefont {Plotkin}}, \ and\ \bibinfo {author} {\bibfnamefont
  {I.}~\bibnamefont {Knezevic}},\ }\href {\doibase
  10.1016/S0140-6736(21)00527-4} {\bibfield  {journal} {\bibinfo  {journal}
  {The Lancet}\ }\textbf {\bibinfo {volume} {397}},\ \bibinfo {pages} {1347}
  (\bibinfo {year} {2021})}\BibitemShut {NoStop}%
\bibitem [{\citenamefont {Knezevic}\ \emph {et~al.}(2022)\citenamefont
  {Knezevic}, \citenamefont {Mattiuzzo}, \citenamefont {Page}, \citenamefont
  {Minor}, \citenamefont {Griffiths}, \citenamefont {Nuebling},\ and\
  \citenamefont {Moorthy}}]{WHO2}%
  \BibitemOpen
  \bibfield  {author} {\bibinfo {author} {\bibfnamefont {I.}~\bibnamefont
  {Knezevic}}, \bibinfo {author} {\bibfnamefont {G.}~\bibnamefont {Mattiuzzo}},
  \bibinfo {author} {\bibfnamefont {M.}~\bibnamefont {Page}}, \bibinfo {author}
  {\bibfnamefont {P.}~\bibnamefont {Minor}}, \bibinfo {author} {\bibfnamefont
  {E.}~\bibnamefont {Griffiths}}, \bibinfo {author} {\bibfnamefont
  {M.}~\bibnamefont {Nuebling}}, \ and\ \bibinfo {author} {\bibfnamefont
  {V.}~\bibnamefont {Moorthy}},\ }\href {\doibase
  10.1016/S2666-5247(21)00266-4} {\bibfield  {journal} {\bibinfo  {journal}
  {The Lancet Microbe}\ }\textbf {\bibinfo {volume} {3}},\ \bibinfo {pages}
  {e235} (\bibinfo {year} {2022})}\BibitemShut {NoStop}%
\bibitem [{Note1()}]{Note1}%
  \BibitemOpen
  \bibinfo {note} {Certain commercial products are referenced (directly or
  indirectly) in this manuscript to clarify our theoretical analysis. Such
  reference does not imply endorsement or approval of any kind by
  NIST.}\BibitemShut {Stop}%
\bibitem [{\citenamefont {Niu}\ \emph {et~al.}(2021)\citenamefont {Niu},
  \citenamefont {Wittrock}, \citenamefont {Clabaugh}, \citenamefont
  {Srivastava},\ and\ \citenamefont {Cho}}]{Binding}%
  \BibitemOpen
  \bibfield  {author} {\bibinfo {author} {\bibfnamefont {L.}~\bibnamefont
  {Niu}}, \bibinfo {author} {\bibfnamefont {K.~N.}\ \bibnamefont {Wittrock}},
  \bibinfo {author} {\bibfnamefont {G.~C.}\ \bibnamefont {Clabaugh}}, \bibinfo
  {author} {\bibfnamefont {V.}~\bibnamefont {Srivastava}}, \ and\ \bibinfo
  {author} {\bibfnamefont {M.~W.}\ \bibnamefont {Cho}},\ }\href {\doibase
  10.3389/fimmu.2021.647934} {\bibfield  {journal} {\bibinfo  {journal}
  {Frontiers in Immunology}\ }\textbf {\bibinfo {volume} {12}} (\bibinfo {year}
  {2021}),\ 10.3389/fimmu.2021.647934}\BibitemShut {NoStop}%
\bibitem [{\citenamefont {Braden}\ and\ \citenamefont
  {Poljak}(1997)}]{Affinity}%
  \BibitemOpen
  \bibfield  {author} {\bibinfo {author} {\bibfnamefont {B.~C.}\ \bibnamefont
  {Braden}}\ and\ \bibinfo {author} {\bibfnamefont {R.~J.}\ \bibnamefont
  {Poljak}},\ }in\ \href {\doibase
  https://doi.org/10.1016/B978-044482807-1/50004-1} {\emph {\bibinfo
  {booktitle} {Idiotypes in Medicine: Autoimmunity, Infection and Cancer}}},\
  \bibinfo {editor} {edited by\ \bibinfo {editor} {\bibfnamefont
  {Y.}~\bibnamefont {Shoenfeld}}, \bibinfo {editor} {\bibfnamefont {R.~C.}\
  \bibnamefont {Kennedy}}, \ and\ \bibinfo {editor} {\bibfnamefont
  {S.}~\bibnamefont {Ferrone}}}\ (\bibinfo  {publisher} {Elsevier Science
  B.V.},\ \bibinfo {address} {Amsterdam},\ \bibinfo {year} {1997})\ pp.\
  \bibinfo {pages} {37--50}\BibitemShut {NoStop}%
\bibitem [{Note2()}]{Note2}%
  \BibitemOpen
  \bibinfo {note} {This mirrors the challenge of estimating binding kinetics
  for emerging SARS-CoV-2 variants.}\BibitemShut {Stop}%
\bibitem [{\citenamefont {Frey}\ \emph {et~al.}(1998)\citenamefont {Frey},
  \citenamefont {{Di Canzio}},\ and\ \citenamefont {Zurakowski}}]{Endpoint}%
  \BibitemOpen
  \bibfield  {author} {\bibinfo {author} {\bibfnamefont {A.}~\bibnamefont
  {Frey}}, \bibinfo {author} {\bibfnamefont {J.}~\bibnamefont {{Di Canzio}}}, \
  and\ \bibinfo {author} {\bibfnamefont {D.}~\bibnamefont {Zurakowski}},\
  }\href {\doibase https://doi.org/10.1016/S0022-1759(98)00170-7} {\bibfield
  {journal} {\bibinfo  {journal} {Journal of Immunological Methods}\ }\textbf
  {\bibinfo {volume} {221}},\ \bibinfo {pages} {35} (\bibinfo {year}
  {1998})}\BibitemShut {NoStop}%
\bibitem [{\citenamefont {Barrette}\ \emph {et~al.}(2006)\citenamefont
  {Barrette}, \citenamefont {Urbonas},\ and\ \citenamefont
  {Silbart}}]{singlepoint}%
  \BibitemOpen
  \bibfield  {author} {\bibinfo {author} {\bibfnamefont {R.~W.}\ \bibnamefont
  {Barrette}}, \bibinfo {author} {\bibfnamefont {J.}~\bibnamefont {Urbonas}}, \
  and\ \bibinfo {author} {\bibfnamefont {L.~K.}\ \bibnamefont {Silbart}},\
  }\href {\doibase 10.1128/CVI.00422-05} {\bibfield  {journal} {\bibinfo
  {journal} {Clinical and Vaccine Immunology}\ }\textbf {\bibinfo {volume}
  {13}},\ \bibinfo {pages} {802} (\bibinfo {year} {2006})}\BibitemShut
  {NoStop}%
\bibitem [{\citenamefont {Wang}\ \emph {et~al.}(2023)\citenamefont {Wang},
  \citenamefont {Patrone}, \citenamefont {Kearsley},\ and\ \citenamefont
  {Lin-Gibson}}]{InterlabLili}%
  \BibitemOpen
  \bibfield  {author} {\bibinfo {author} {\bibfnamefont {L.}~\bibnamefont
  {Wang}}, \bibinfo {author} {\bibfnamefont {P.~N.}\ \bibnamefont {Patrone}},
  \bibinfo {author} {\bibfnamefont {A.~J.}\ \bibnamefont {Kearsley}}, \ and\
  \bibinfo {author} {\bibfnamefont {S.}~\bibnamefont {Lin-Gibson}},\
  }\href@noop {} {\bibfield  {journal} {\bibinfo  {journal} {Submitted}\ }
  (\bibinfo {year} {2023})}\BibitemShut {NoStop}%
\bibitem [{\citenamefont {Smith}(2013)}]{SmithUQ}%
  \BibitemOpen
  \bibfield  {author} {\bibinfo {author} {\bibfnamefont {R.}~\bibnamefont
  {Smith}},\ }\href@noop {} {\emph {\bibinfo {title} {Uncertainty
  Quantification: Theory, Implementation, and Applications}}},\ Computational
  Science and Engineering\ (\bibinfo  {publisher} {Society for Industrial and
  Applied Mathematics},\ \bibinfo {year} {2013})\BibitemShut {NoStop}%
\bibitem [{\citenamefont {Tate}\ \emph {et~al.}(2012)\citenamefont {Tate},
  \citenamefont {Johnson},\ and\ \citenamefont {Legg}}]{harmonization1}%
  \BibitemOpen
  \bibfield  {author} {\bibinfo {author} {\bibfnamefont {J.~R.}\ \bibnamefont
  {Tate}}, \bibinfo {author} {\bibfnamefont {R.}~\bibnamefont {Johnson}}, \
  and\ \bibinfo {author} {\bibfnamefont {M.}~\bibnamefont {Legg}},\ }\href@noop
  {} {\bibfield  {journal} {\bibinfo  {journal} {Clin Biochem Rev}\ }\textbf
  {\bibinfo {volume} {33}},\ \bibinfo {pages} {81} (\bibinfo {year}
  {2012})}\BibitemShut {NoStop}%
\bibitem [{\citenamefont {McLawhon}(2011)}]{harmonization2}%
  \BibitemOpen
  \bibfield  {author} {\bibinfo {author} {\bibfnamefont {R.~W.}\ \bibnamefont
  {McLawhon}},\ }\href {\doibase 10.1373/clinchem.2011.166041} {\bibfield
  {journal} {\bibinfo  {journal} {Clinical Chemistry}\ }\textbf {\bibinfo
  {volume} {57}},\ \bibinfo {pages} {936} (\bibinfo {year} {2011})},\ \Eprint
  {http://arxiv.org/abs/https://academic.oup.com/clinchem/article-pdf/57/7/936/32655704/clinchem0936.pdf}
  {https://academic.oup.com/clinchem/article-pdf/57/7/936/32655704/clinchem0936.pdf}
  \BibitemShut {NoStop}%
\bibitem [{\citenamefont {Kemp}\ \emph {et~al.}(2022)\citenamefont {Kemp},
  \citenamefont {Quesinberry}, \citenamefont {Cherry}, \citenamefont {Lowy},\
  and\ \citenamefont {Pinto}}]{NCI}%
  \BibitemOpen
  \bibfield  {author} {\bibinfo {author} {\bibfnamefont {T.~J.}\ \bibnamefont
  {Kemp}}, \bibinfo {author} {\bibfnamefont {J.~T.}\ \bibnamefont
  {Quesinberry}}, \bibinfo {author} {\bibfnamefont {J.}~\bibnamefont {Cherry}},
  \bibinfo {author} {\bibfnamefont {D.~R.}\ \bibnamefont {Lowy}}, \ and\
  \bibinfo {author} {\bibfnamefont {L.~A.}\ \bibnamefont {Pinto}},\ }\href
  {\doibase 10.1128/jcm.00995-22} {\bibfield  {journal} {\bibinfo  {journal}
  {Journal of Clinical Microbiology}\ }\textbf {\bibinfo {volume} {60}},\
  \bibinfo {pages} {e00995} (\bibinfo {year} {2022})}\BibitemShut {NoStop}%
\bibitem [{\citenamefont {Bentley}\ \emph {et~al.}(2022)\citenamefont
  {Bentley}, \citenamefont {Atkinson}, \citenamefont {Rigsby}, \citenamefont
  {Elsley}, \citenamefont {Bernasconi}, \citenamefont {Kristiansen},
  \citenamefont {Harvala}, \citenamefont {Turtle}, \citenamefont {Dobson},
  \citenamefont {Wendel}, \citenamefont {Anderson}, \citenamefont {Kempster},
  \citenamefont {Duran}, \citenamefont {Padley}, \citenamefont {Almond},
  \citenamefont {Rose}, \citenamefont {Page},\ and\ \citenamefont
  {Mattiuzzo}}]{WHO3}%
  \BibitemOpen
  \bibfield  {author} {\bibinfo {author} {\bibfnamefont {E.~M.}\ \bibnamefont
  {Bentley}}, \bibinfo {author} {\bibfnamefont {E.}~\bibnamefont {Atkinson}},
  \bibinfo {author} {\bibfnamefont {P.}~\bibnamefont {Rigsby}}, \bibinfo
  {author} {\bibfnamefont {W.}~\bibnamefont {Elsley}}, \bibinfo {author}
  {\bibfnamefont {V.}~\bibnamefont {Bernasconi}}, \bibinfo {author}
  {\bibfnamefont {P.}~\bibnamefont {Kristiansen}}, \bibinfo {author}
  {\bibfnamefont {H.}~\bibnamefont {Harvala}}, \bibinfo {author} {\bibfnamefont
  {L.~C.}\ \bibnamefont {Turtle}}, \bibinfo {author} {\bibfnamefont
  {S.}~\bibnamefont {Dobson}}, \bibinfo {author} {\bibfnamefont
  {S.}~\bibnamefont {Wendel}}, \bibinfo {author} {\bibfnamefont
  {R.}~\bibnamefont {Anderson}}, \bibinfo {author} {\bibfnamefont
  {S.}~\bibnamefont {Kempster}}, \bibinfo {author} {\bibfnamefont
  {J.}~\bibnamefont {Duran}}, \bibinfo {author} {\bibfnamefont
  {D.}~\bibnamefont {Padley}}, \bibinfo {author} {\bibfnamefont
  {N.}~\bibnamefont {Almond}}, \bibinfo {author} {\bibfnamefont {N.~J.}\
  \bibnamefont {Rose}}, \bibinfo {author} {\bibfnamefont {M.}~\bibnamefont
  {Page}}, \ and\ \bibinfo {author} {\bibfnamefont {G.}~\bibnamefont
  {Mattiuzzo}},\ }\href@noop {} {\bibfield  {journal} {\bibinfo  {journal}
  {Expert Committee on Biological Standardization}\ ,\ \bibinfo {pages} {1}}
  (\bibinfo {year} {2022})}\BibitemShut {NoStop}%
\bibitem [{\citenamefont {Nixon}\ \emph {et~al.}(2022)\citenamefont {Nixon},
  \citenamefont {Jindal}, \citenamefont {Parker}, \citenamefont {Reich},
  \citenamefont {Ghobadi}, \citenamefont {Lee}, \citenamefont {Truelove},\ and\
  \citenamefont {Gardner}}]{COVIDUQ1}%
  \BibitemOpen
  \bibfield  {author} {\bibinfo {author} {\bibfnamefont {K.}~\bibnamefont
  {Nixon}}, \bibinfo {author} {\bibfnamefont {S.}~\bibnamefont {Jindal}},
  \bibinfo {author} {\bibfnamefont {F.}~\bibnamefont {Parker}}, \bibinfo
  {author} {\bibfnamefont {N.~G.}\ \bibnamefont {Reich}}, \bibinfo {author}
  {\bibfnamefont {K.}~\bibnamefont {Ghobadi}}, \bibinfo {author} {\bibfnamefont
  {E.~C.}\ \bibnamefont {Lee}}, \bibinfo {author} {\bibfnamefont
  {S.}~\bibnamefont {Truelove}}, \ and\ \bibinfo {author} {\bibfnamefont
  {L.}~\bibnamefont {Gardner}},\ }\href {\doibase
  10.1016/S2589-7500(22)00148-0} {\bibfield  {journal} {\bibinfo  {journal}
  {The Lancet Digital Health}\ }\textbf {\bibinfo {volume} {4}},\ \bibinfo
  {pages} {e738} (\bibinfo {year} {2022})}\BibitemShut {NoStop}%
\bibitem [{\citenamefont {Dron}\ \emph {et~al.}(2022)\citenamefont {Dron},
  \citenamefont {Kalatharan}, \citenamefont {Gupta}, \citenamefont {Haggstrom},
  \citenamefont {Zariffa}, \citenamefont {Morris}, \citenamefont {Arora},\ and\
  \citenamefont {Park}}]{COVIDUQ2}%
  \BibitemOpen
  \bibfield  {author} {\bibinfo {author} {\bibfnamefont {L.}~\bibnamefont
  {Dron}}, \bibinfo {author} {\bibfnamefont {V.}~\bibnamefont {Kalatharan}},
  \bibinfo {author} {\bibfnamefont {A.}~\bibnamefont {Gupta}}, \bibinfo
  {author} {\bibfnamefont {J.}~\bibnamefont {Haggstrom}}, \bibinfo {author}
  {\bibfnamefont {N.}~\bibnamefont {Zariffa}}, \bibinfo {author} {\bibfnamefont
  {A.~D.}\ \bibnamefont {Morris}}, \bibinfo {author} {\bibfnamefont
  {P.}~\bibnamefont {Arora}}, \ and\ \bibinfo {author} {\bibfnamefont
  {J.}~\bibnamefont {Park}},\ }\href {\doibase 10.1016/S2589-7500(22)00147-9}
  {\bibfield  {journal} {\bibinfo  {journal} {The Lancet Digital Health}\
  }\textbf {\bibinfo {volume} {4}},\ \bibinfo {pages} {e748} (\bibinfo {year}
  {2022})}\BibitemShut {NoStop}%
\bibitem [{Note3()}]{Note3}%
  \BibitemOpen
  \bibinfo {note} {We do not assume that normalization and harmonization are
  necessarily identical. In the next section, however, we determine conditions
  under which these tasks are the same.}\BibitemShut {Stop}%
\bibitem [{Note4()}]{Note4}%
  \BibitemOpen
  \bibinfo {note} {In later sections we find it more convenient to dispense
  with this linearity assumption.}\BibitemShut {Stop}%
\bibitem [{Note5()}]{Note5}%
  \BibitemOpen
  \bibinfo {note} {When $y_r$ is unknown, as is the case for human-derived,
  pooled samples, its value is assigned arbitrarily \cite {WHO1,WHO2,NCI}. In
  this case, it is equally valid to fix $c_r$ instead. This is purely a
  semantic choice that does not play a role in our analysis.}\BibitemShut
  {Stop}%
\bibitem [{\citenamefont {Holland}\ and\ \citenamefont
  {Holland}(2004)}]{massaction}%
  \BibitemOpen
  \bibfield  {author} {\bibinfo {author} {\bibfnamefont {T.}~\bibnamefont
  {Holland}}\ and\ \bibinfo {author} {\bibfnamefont {H.}~\bibnamefont
  {Holland}},\ }\href {\doibase
  https://doi.org/10.1111/j.0022-2720.2004.01294.x} {\bibfield  {journal}
  {\bibinfo  {journal} {Journal of Microscopy}\ }\textbf {\bibinfo {volume}
  {214}},\ \bibinfo {pages} {1} (\bibinfo {year} {2004})}\BibitemShut {NoStop}%
\bibitem [{\citenamefont {Božič}\ \emph {et~al.}(2014)\citenamefont
  {Božič}, \citenamefont {Čučnik}, \citenamefont {Kveder},\ and\
  \citenamefont {Rozman}}]{Affinity2}%
  \BibitemOpen
  \bibfield  {author} {\bibinfo {author} {\bibfnamefont {B.}~\bibnamefont
  {Božič}}, \bibinfo {author} {\bibfnamefont {S.}~\bibnamefont {Čučnik}},
  \bibinfo {author} {\bibfnamefont {T.}~\bibnamefont {Kveder}}, \ and\ \bibinfo
  {author} {\bibfnamefont {B.}~\bibnamefont {Rozman}},\ }in\ \href {\doibase
  https://doi.org/10.1016/B978-0-444-56378-1.00005-8} {\emph {\bibinfo
  {booktitle} {Autoantibodies (Third Edition)}}},\ \bibinfo {editor} {edited
  by\ \bibinfo {editor} {\bibfnamefont {Y.}~\bibnamefont {Shoenfeld}}, \bibinfo
  {editor} {\bibfnamefont {P.~L.}\ \bibnamefont {Meroni}}, \ and\ \bibinfo
  {editor} {\bibfnamefont {M.~E.}\ \bibnamefont {Gershwin}}}\ (\bibinfo
  {publisher} {Elsevier},\ \bibinfo {address} {San Diego},\ \bibinfo {year}
  {2014})\ \bibinfo {edition} {third edition}\ ed.,\ pp.\ \bibinfo {pages}
  {43--49}\BibitemShut {NoStop}%
\bibitem [{\citenamefont {Pathria}(2016)}]{Pathria}%
  \BibitemOpen
  \bibfield  {author} {\bibinfo {author} {\bibfnamefont {R.}~\bibnamefont
  {Pathria}},\ }\href@noop {} {\emph {\bibinfo {title} {Statistical
  Mechanics}}}\ (\bibinfo  {publisher} {Elsevier Science},\ \bibinfo {year}
  {2016})\BibitemShut {NoStop}%
\bibitem [{\citenamefont {Erickson}\ and\ \citenamefont
  {Corbin~Goodman}(2022)}]{Monovalent}%
  \BibitemOpen
  \bibfield  {author} {\bibinfo {author} {\bibfnamefont {H.~P.}\ \bibnamefont
  {Erickson}}\ and\ \bibinfo {author} {\bibfnamefont {L.}~\bibnamefont
  {Corbin~Goodman}},\ }\href {\doibase 10.1021/acs.biochem.2c00291} {\bibfield
  {journal} {\bibinfo  {journal} {Biochemistry}\ } (\bibinfo {year} {2022}),\
  10.1021/acs.biochem.2c00291}\BibitemShut {NoStop}%
\bibitem [{\citenamefont {Andreasson}\ \emph {et~al.}(2015)\citenamefont
  {Andreasson}, \citenamefont {Perret-Liaudet}, \citenamefont {van Waalwijk~van
  Doorn}, \citenamefont {Blennow}, \citenamefont {Chiasserini}, \citenamefont
  {Engelborghs}, \citenamefont {Fladby}, \citenamefont {Genc}, \citenamefont
  {Kruse}, \citenamefont {Kuiperij}, \citenamefont {Kulic}, \citenamefont
  {Lewczuk}, \citenamefont {Mollenhauer}, \citenamefont {Mroczko},
  \citenamefont {Parnetti}, \citenamefont {Vanmechelen}, \citenamefont
  {Verbeek}, \citenamefont {Winblad}, \citenamefont {Zetterberg}, \citenamefont
  {Koel-Simmelink},\ and\ \citenamefont {Teunissen}}]{linearity}%
  \BibitemOpen
  \bibfield  {author} {\bibinfo {author} {\bibfnamefont {U.}~\bibnamefont
  {Andreasson}}, \bibinfo {author} {\bibfnamefont {A.}~\bibnamefont
  {Perret-Liaudet}}, \bibinfo {author} {\bibfnamefont {L.~J.~C.}\ \bibnamefont
  {van Waalwijk~van Doorn}}, \bibinfo {author} {\bibfnamefont {K.}~\bibnamefont
  {Blennow}}, \bibinfo {author} {\bibfnamefont {D.}~\bibnamefont
  {Chiasserini}}, \bibinfo {author} {\bibfnamefont {S.}~\bibnamefont
  {Engelborghs}}, \bibinfo {author} {\bibfnamefont {T.}~\bibnamefont {Fladby}},
  \bibinfo {author} {\bibfnamefont {S.}~\bibnamefont {Genc}}, \bibinfo {author}
  {\bibfnamefont {N.}~\bibnamefont {Kruse}}, \bibinfo {author} {\bibfnamefont
  {H.~B.}\ \bibnamefont {Kuiperij}}, \bibinfo {author} {\bibfnamefont
  {L.}~\bibnamefont {Kulic}}, \bibinfo {author} {\bibfnamefont
  {P.}~\bibnamefont {Lewczuk}}, \bibinfo {author} {\bibfnamefont
  {B.}~\bibnamefont {Mollenhauer}}, \bibinfo {author} {\bibfnamefont
  {B.}~\bibnamefont {Mroczko}}, \bibinfo {author} {\bibfnamefont
  {L.}~\bibnamefont {Parnetti}}, \bibinfo {author} {\bibfnamefont
  {E.}~\bibnamefont {Vanmechelen}}, \bibinfo {author} {\bibfnamefont {M.~M.}\
  \bibnamefont {Verbeek}}, \bibinfo {author} {\bibfnamefont {B.}~\bibnamefont
  {Winblad}}, \bibinfo {author} {\bibfnamefont {H.}~\bibnamefont {Zetterberg}},
  \bibinfo {author} {\bibfnamefont {M.}~\bibnamefont {Koel-Simmelink}}, \ and\
  \bibinfo {author} {\bibfnamefont {C.~E.}\ \bibnamefont {Teunissen}},\ }\href
  {\doibase 10.3389/fneur.2015.00179} {\bibfield  {journal} {\bibinfo
  {journal} {Frontiers in Neurology}\ }\textbf {\bibinfo {volume} {6}}
  (\bibinfo {year} {2015}),\ 10.3389/fneur.2015.00179}\BibitemShut {NoStop}%
\bibitem [{Note6()}]{Note6}%
  \BibitemOpen
  \bibinfo {note} {It is necessary to distinguish nonlinearity in the
  measurement due to Eq.\ \protect \eqref {eq:equilibrium} from nonlinear
  effects due to detection equipment such as photodetectors. This distinction
  is important for the analysis in Sec.\ \ref {sec:affine}.}\BibitemShut
  {Stop}%
\bibitem [{Note7()}]{Note7}%
  \BibitemOpen
  \bibinfo {note} {\textcolor {black}{When $\protect \hat K_{m,n} \ll 1$ unit
  volume, $K_{m,n}=\protect \hat K_{m,n}b_n$. When $b_n \gg y_m$, one finds
  $K_{m,n}=1$.}}\BibitemShut {Stop}%
\bibitem [{Note8()}]{Note8}%
  \BibitemOpen
  \bibinfo {note} {Note that the $\kappa _{{\protect \rm Y}_m}$ and $\kappa
  _{{\protect \rm B}_n}$ are not determined uniquely by Eq.\ \protect \eqref
  {eq:separability}. We can always define new constants $\protect \hat \kappa
  _{{\protect \rm Y}_m}= \kappa _{{\protect \rm Y}_m}/\alpha $ and $\protect
  \hat \kappa _{{\protect \rm B}_n}=\kappa _{{\protect \rm B}_n}\alpha $ for
  any positive constant $\alpha $ such that the product $K_{m,n}=\protect \hat
  \kappa _{{\protect \rm Y}_m}\protect \hat \kappa _{{\protect \rm B}_n} =
  \kappa _{{\protect \rm Y}_m} \kappa _{{\protect \rm B}_n}$ is
  unchanged.}\BibitemShut {Stop}%
\bibitem [{Note9()}]{Note9}%
  \BibitemOpen
  \bibinfo {note} {We do not explore the theoretical question of whether
  separability is {\protect \it necessary} for harmonization.}\BibitemShut
  {Stop}%
\bibitem [{Note10()}]{Note10}%
  \BibitemOpen
  \bibinfo {note} {\textcolor {black}{More precisely, when viewed as a matrix
  with elements $s,n$, we require $K_{s,n}$ to have rank greater than
  one.}}\BibitemShut {Stop}%
\bibitem [{\citenamefont {Rukhin}(2009)}]{Rukhin}%
  \BibitemOpen
  \bibfield  {author} {\bibinfo {author} {\bibfnamefont {A.~L.}\ \bibnamefont
  {Rukhin}},\ }\href {\doibase 10.1088/0026-1394/46/3/021} {\bibfield
  {journal} {\bibinfo  {journal} {Metrologia}\ }\textbf {\bibinfo {volume}
  {46}},\ \bibinfo {pages} {323} (\bibinfo {year} {2009})}\BibitemShut
  {NoStop}%
\bibitem [{\citenamefont {Rasmussen}\ and\ \citenamefont
  {Williams}(2005)}]{RW}%
  \BibitemOpen
  \bibfield  {author} {\bibinfo {author} {\bibfnamefont {C.~E.}\ \bibnamefont
  {Rasmussen}}\ and\ \bibinfo {author} {\bibfnamefont {C.~K.~I.}\ \bibnamefont
  {Williams}},\ }\href {\doibase 10.7551/mitpress/3206.001.0001} {\emph
  {\bibinfo {title} {{Gaussian Processes for Machine Learning}}}}\ (\bibinfo
  {publisher} {The MIT Press},\ \bibinfo {year} {2005})\BibitemShut {NoStop}%
\bibitem [{Note11()}]{Note11}%
  \BibitemOpen
  \bibinfo {note} {The parameter $d$ can of course be greater than unity, but
  this is interpreted as a sample being made more concentrated, which is not
  typical in experimental settings.}\BibitemShut {Stop}%
\bibitem [{\citenamefont {Patrone}\ \emph
  {et~al.}(2020{\natexlab{a}})\citenamefont {Patrone}, \citenamefont {Romsos},
  \citenamefont {Cleveland}, \citenamefont {Vallone},\ and\ \citenamefont
  {Kearsley}}]{qPCR}%
  \BibitemOpen
  \bibfield  {author} {\bibinfo {author} {\bibfnamefont {P.~N.}\ \bibnamefont
  {Patrone}}, \bibinfo {author} {\bibfnamefont {E.~L.}\ \bibnamefont {Romsos}},
  \bibinfo {author} {\bibfnamefont {M.~H.}\ \bibnamefont {Cleveland}}, \bibinfo
  {author} {\bibfnamefont {P.~M.}\ \bibnamefont {Vallone}}, \ and\ \bibinfo
  {author} {\bibfnamefont {A.~J.}\ \bibnamefont {Kearsley}},\ }\href {\doibase
  10.1007/s00216-020-02930-z} {\bibfield  {journal} {\bibinfo  {journal}
  {Analytical and Bioanalytical Chemistry}\ }\textbf {\bibinfo {volume}
  {412}},\ \bibinfo {pages} {7977} (\bibinfo {year}
  {2020}{\natexlab{a}})}\BibitemShut {NoStop}%
\bibitem [{\citenamefont {Patrone}\ \emph
  {et~al.}(2020{\natexlab{b}})\citenamefont {Patrone}, \citenamefont
  {Kearsley}, \citenamefont {Majikes},\ and\ \citenamefont {Liddle}}]{PCR2}%
  \BibitemOpen
  \bibfield  {author} {\bibinfo {author} {\bibfnamefont {P.~N.}\ \bibnamefont
  {Patrone}}, \bibinfo {author} {\bibfnamefont {A.~J.}\ \bibnamefont
  {Kearsley}}, \bibinfo {author} {\bibfnamefont {J.~M.}\ \bibnamefont
  {Majikes}}, \ and\ \bibinfo {author} {\bibfnamefont {J.~A.}\ \bibnamefont
  {Liddle}},\ }\href {\doibase https://doi.org/10.1016/j.ab.2020.113773}
  {\bibfield  {journal} {\bibinfo  {journal} {Analytical Biochemistry}\
  }\textbf {\bibinfo {volume} {607}},\ \bibinfo {pages} {113773} (\bibinfo
  {year} {2020}{\natexlab{b}})}\BibitemShut {NoStop}%
\bibitem [{\citenamefont {Patrone}\ \emph {et~al.}(2019)\citenamefont
  {Patrone}, \citenamefont {Cooksey},\ and\ \citenamefont
  {Kearsley}}]{flowmeter}%
  \BibitemOpen
  \bibfield  {author} {\bibinfo {author} {\bibfnamefont {P.~N.}\ \bibnamefont
  {Patrone}}, \bibinfo {author} {\bibfnamefont {G.}~\bibnamefont {Cooksey}}, \
  and\ \bibinfo {author} {\bibfnamefont {A.}~\bibnamefont {Kearsley}},\ }\href
  {\doibase 10.1103/PhysRevApplied.11.034025} {\bibfield  {journal} {\bibinfo
  {journal} {Phys. Rev. Applied}\ }\textbf {\bibinfo {volume} {11}},\ \bibinfo
  {pages} {034025} (\bibinfo {year} {2019})}\BibitemShut {NoStop}%
\bibitem [{\citenamefont {STINE}(1989)}]{Bootstrap1}%
  \BibitemOpen
  \bibfield  {author} {\bibinfo {author} {\bibfnamefont {R.}~\bibnamefont
  {STINE}},\ }\href {\doibase 10.1177/0049124189018002003} {\bibfield
  {journal} {\bibinfo  {journal} {Sociological Methods \& Research}\ }\textbf
  {\bibinfo {volume} {18}},\ \bibinfo {pages} {243} (\bibinfo {year}
  {1989})}\BibitemShut {NoStop}%
\bibitem [{\citenamefont {Chernick}\ \emph {et~al.}(2011)\citenamefont
  {Chernick}, \citenamefont {Gonz{\'a}lez-Manteiga}, \citenamefont
  {Crujeiras},\ and\ \citenamefont {Barrios}}]{Bootstrap2}%
  \BibitemOpen
  \bibfield  {author} {\bibinfo {author} {\bibfnamefont {M.~R.}\ \bibnamefont
  {Chernick}}, \bibinfo {author} {\bibfnamefont {W.}~\bibnamefont
  {Gonz{\'a}lez-Manteiga}}, \bibinfo {author} {\bibfnamefont {R.~M.}\
  \bibnamefont {Crujeiras}}, \ and\ \bibinfo {author} {\bibfnamefont {E.~B.}\
  \bibnamefont {Barrios}},\ }\enquote {\bibinfo {title} {Bootstrap methods},}\
  in\ \href {\doibase 10.1007/978-3-642-04898-2_150} {\emph {\bibinfo
  {booktitle} {International Encyclopedia of Statistical Science}}},\ \bibinfo
  {editor} {edited by\ \bibinfo {editor} {\bibfnamefont {M.}~\bibnamefont
  {Lovric}}}\ (\bibinfo  {publisher} {Springer Berlin Heidelberg},\ \bibinfo
  {address} {Berlin, Heidelberg},\ \bibinfo {year} {2011})\ pp.\ \bibinfo
  {pages} {169--174}\BibitemShut {NoStop}%
\bibitem [{\citenamefont {BIPM}\ \emph {et~al.}()\citenamefont {BIPM},
  \citenamefont {IEC}, \citenamefont {IFCC}, \citenamefont {ILAC},
  \citenamefont {ISO}, \citenamefont {IUPAC}, \citenamefont {IUPAP},\ and\
  \citenamefont {OIML}}]{GUM}%
  \BibitemOpen
  \bibfield  {author} {\bibinfo {author} {\bibnamefont {BIPM}}, \bibinfo
  {author} {\bibnamefont {IEC}}, \bibinfo {author} {\bibnamefont {IFCC}},
  \bibinfo {author} {\bibnamefont {ILAC}}, \bibinfo {author} {\bibnamefont
  {ISO}}, \bibinfo {author} {\bibnamefont {IUPAC}}, \bibinfo {author}
  {\bibnamefont {IUPAP}}, \ and\ \bibinfo {author} {\bibnamefont {OIML}},\
  }\href
  {https://www.bipm.org/documents/20126/2071204/JCGM\_100\_2008\_E.pdf/cb0ef43f-baa5-11cf-3f85-4dcd86f77bd6}
  {\enquote {\bibinfo {title} {Evaluation of measurement data --- {G}uide to
  the expression of uncertainty in measurement},}\ }\bibinfo {howpublished}
  {Joint Committee for Guides in Metrology, JCGM 100:2008}\BibitemShut
  {NoStop}%
\bibitem [{Note12()}]{Note12}%
  \BibitemOpen
  \bibinfo {note} {Recall that the logarithm of concentration is linear in the
  Gibbs free energy. Thus Eq.\ \protect \eqref {eq:logavg} can also be viewed
  as an estimate of the average $\Delta G$.}\BibitemShut {Stop}%
\bibitem [{\citenamefont {Clifford~Cohen}(2016)}]{censor1}%
  \BibitemOpen
  \bibfield  {author} {\bibinfo {author} {\bibfnamefont {A.}~\bibnamefont
  {Clifford~Cohen}},\ }\href@noop {} {\emph {\bibinfo {title} {Truncated and
  censored samples}}}\ (\bibinfo  {publisher} {CRC Press},\ \bibinfo {year}
  {2016})\BibitemShut {NoStop}%
\bibitem [{\citenamefont {Tony}(2011)}]{censor2}%
  \BibitemOpen
  \bibfield  {author} {\bibinfo {author} {\bibfnamefont {N.~H.~K.}\
  \bibnamefont {Tony}},\ }\enquote {\bibinfo {title} {Censoring methodology},}\
  in\ \href@noop {} {\emph {\bibinfo {booktitle} {International Encyclopedia of
  Statistical Science}}},\ \bibinfo {editor} {edited by\ \bibinfo {editor}
  {\bibfnamefont {M.}~\bibnamefont {Lovric}}}\ (\bibinfo  {publisher} {Springer
  Berlin Heidelberg},\ \bibinfo {address} {Berlin, Heidelberg},\ \bibinfo
  {year} {2011})\ pp.\ \bibinfo {pages} {221--224}\BibitemShut {NoStop}%
\bibitem [{\citenamefont {Muller}\ \emph {et~al.}(2022)\citenamefont {Muller},
  \citenamefont {Kannenberg}, \citenamefont {Biemann}, \citenamefont
  {Hönemann}, \citenamefont {Ackermann},\ and\ \citenamefont
  {Jassoy}}]{Harmon1}%
  \BibitemOpen
  \bibfield  {author} {\bibinfo {author} {\bibfnamefont {L.}~\bibnamefont
  {Muller}}, \bibinfo {author} {\bibfnamefont {J.}~\bibnamefont {Kannenberg}},
  \bibinfo {author} {\bibfnamefont {R.}~\bibnamefont {Biemann}}, \bibinfo
  {author} {\bibfnamefont {M.}~\bibnamefont {Hönemann}}, \bibinfo {author}
  {\bibfnamefont {G.}~\bibnamefont {Ackermann}}, \ and\ \bibinfo {author}
  {\bibfnamefont {C.}~\bibnamefont {Jassoy}},\ }\href {\doibase
  https://doi.org/10.1016/j.jcv.2022.105269} {\bibfield  {journal} {\bibinfo
  {journal} {Journal of Clinical Virology}\ }\textbf {\bibinfo {volume}
  {155}},\ \bibinfo {pages} {105269} (\bibinfo {year} {2022})}\BibitemShut
  {NoStop}%
\bibitem [{\citenamefont {Perkmann}\ \emph {et~al.}(2021)\citenamefont
  {Perkmann}, \citenamefont {Perkmann-Nagele}, \citenamefont {Koller},
  \citenamefont {Mucher}, \citenamefont {Radakovics}, \citenamefont
  {Marculescu}, \citenamefont {Wolzt}, \citenamefont {Wagner}, \citenamefont
  {Binder},\ and\ \citenamefont {Haslacher}}]{Harmon2}%
  \BibitemOpen
  \bibfield  {author} {\bibinfo {author} {\bibfnamefont {T.}~\bibnamefont
  {Perkmann}}, \bibinfo {author} {\bibfnamefont {N.}~\bibnamefont
  {Perkmann-Nagele}}, \bibinfo {author} {\bibfnamefont {T.}~\bibnamefont
  {Koller}}, \bibinfo {author} {\bibfnamefont {P.}~\bibnamefont {Mucher}},
  \bibinfo {author} {\bibfnamefont {A.}~\bibnamefont {Radakovics}}, \bibinfo
  {author} {\bibfnamefont {R.}~\bibnamefont {Marculescu}}, \bibinfo {author}
  {\bibfnamefont {M.}~\bibnamefont {Wolzt}}, \bibinfo {author} {\bibfnamefont
  {O.~F.}\ \bibnamefont {Wagner}}, \bibinfo {author} {\bibfnamefont {C.~J.}\
  \bibnamefont {Binder}}, \ and\ \bibinfo {author} {\bibfnamefont
  {H.}~\bibnamefont {Haslacher}},\ }\href {\doibase 10.1128/Spectrum.00247-21}
  {\bibfield  {journal} {\bibinfo  {journal} {Microbiology Spectrum}\ }\textbf
  {\bibinfo {volume} {9}},\ \bibinfo {pages} {e00247} (\bibinfo {year}
  {2021})}\BibitemShut {NoStop}%
\bibitem [{\citenamefont {Giavarina}\ and\ \citenamefont
  {Carta}(2022)}]{Harmon3}%
  \BibitemOpen
  \bibfield  {author} {\bibinfo {author} {\bibfnamefont {D.}~\bibnamefont
  {Giavarina}}\ and\ \bibinfo {author} {\bibfnamefont {M.}~\bibnamefont
  {Carta}},\ }\href {\doibase doi:10.1515/dx-2021-0126} {\bibfield  {journal}
  {\bibinfo  {journal} {Diagnosis}\ }\textbf {\bibinfo {volume} {9}},\ \bibinfo
  {pages} {274} (\bibinfo {year} {2022})}\BibitemShut {NoStop}%
\bibitem [{\citenamefont {Infantino}\ \emph {et~al.}(2021)\citenamefont
  {Infantino}, \citenamefont {Pieri}, \citenamefont {Nuccetelli}, \citenamefont
  {Grossi}, \citenamefont {Lari}, \citenamefont {Tomassetti}, \citenamefont
  {Calugi}, \citenamefont {Pancani}, \citenamefont {Benucci}, \citenamefont
  {Casprini}, \citenamefont {Manfredi},\ and\ \citenamefont
  {Bernardini}}]{Harmon4}%
  \BibitemOpen
  \bibfield  {author} {\bibinfo {author} {\bibfnamefont {M.}~\bibnamefont
  {Infantino}}, \bibinfo {author} {\bibfnamefont {M.}~\bibnamefont {Pieri}},
  \bibinfo {author} {\bibfnamefont {M.}~\bibnamefont {Nuccetelli}}, \bibinfo
  {author} {\bibfnamefont {V.}~\bibnamefont {Grossi}}, \bibinfo {author}
  {\bibfnamefont {B.}~\bibnamefont {Lari}}, \bibinfo {author} {\bibfnamefont
  {F.}~\bibnamefont {Tomassetti}}, \bibinfo {author} {\bibfnamefont
  {G.}~\bibnamefont {Calugi}}, \bibinfo {author} {\bibfnamefont
  {S.}~\bibnamefont {Pancani}}, \bibinfo {author} {\bibfnamefont
  {M.}~\bibnamefont {Benucci}}, \bibinfo {author} {\bibfnamefont
  {P.}~\bibnamefont {Casprini}}, \bibinfo {author} {\bibfnamefont
  {M.}~\bibnamefont {Manfredi}}, \ and\ \bibinfo {author} {\bibfnamefont
  {S.}~\bibnamefont {Bernardini}},\ }\href {\doibase
  https://doi.org/10.1016/j.intimp.2021.108095} {\bibfield  {journal} {\bibinfo
   {journal} {International Immunopharmacology}\ }\textbf {\bibinfo {volume}
  {100}},\ \bibinfo {pages} {108095} (\bibinfo {year} {2021})}\BibitemShut
  {NoStop}%
\bibitem [{Note13()}]{Note13}%
  \BibitemOpen
  \bibinfo {note} {In contrast, using different study designs and analyses,
  Refs.\ \cite {Harmon1,Harmon2,Harmon3,Harmon4} found that normalization still
  yielded up to 20-fold systematic discrepancies between assays . Comparing
  these results to Ref.\ \cite {NCI} is challenging due to differences in the
  way results were reported. Reference \cite {NCI} provided aggregate
  coefficients of variation across assays, whereas the other studies explicitly
  correlated normalized results between assays head-to-head.}\BibitemShut
  {Stop}%
\bibitem [{\citenamefont {Haddad}\ \emph {et~al.}(2017)\citenamefont {Haddad},
  \citenamefont {Seifert}, \citenamefont {Chao}, \citenamefont {Possolo},
  \citenamefont {Newell}, \citenamefont {Pratt}, \citenamefont {Williams},\
  and\ \citenamefont {Schlamminger}}]{Planck1}%
  \BibitemOpen
  \bibfield  {author} {\bibinfo {author} {\bibfnamefont {D.}~\bibnamefont
  {Haddad}}, \bibinfo {author} {\bibfnamefont {F.}~\bibnamefont {Seifert}},
  \bibinfo {author} {\bibfnamefont {L.~S.}\ \bibnamefont {Chao}}, \bibinfo
  {author} {\bibfnamefont {A.}~\bibnamefont {Possolo}}, \bibinfo {author}
  {\bibfnamefont {D.~B.}\ \bibnamefont {Newell}}, \bibinfo {author}
  {\bibfnamefont {J.~R.}\ \bibnamefont {Pratt}}, \bibinfo {author}
  {\bibfnamefont {C.~J.}\ \bibnamefont {Williams}}, \ and\ \bibinfo {author}
  {\bibfnamefont {S.}~\bibnamefont {Schlamminger}},\ }\href {\doibase
  10.1088/1681-7575/aa7bf2} {\bibfield  {journal} {\bibinfo  {journal}
  {Metrologia}\ }\textbf {\bibinfo {volume} {54}},\ \bibinfo {pages} {633}
  (\bibinfo {year} {2017})}\BibitemShut {NoStop}%
\bibitem [{\citenamefont {Fang}\ \emph {et~al.}(2020)\citenamefont {Fang},
  \citenamefont {Bielsa}, \citenamefont {Li}, \citenamefont {Kiss},\ and\
  \citenamefont {Stock}}]{Planck2}%
  \BibitemOpen
  \bibfield  {author} {\bibinfo {author} {\bibfnamefont {H.}~\bibnamefont
  {Fang}}, \bibinfo {author} {\bibfnamefont {F.}~\bibnamefont {Bielsa}},
  \bibinfo {author} {\bibfnamefont {S.}~\bibnamefont {Li}}, \bibinfo {author}
  {\bibfnamefont {A.}~\bibnamefont {Kiss}}, \ and\ \bibinfo {author}
  {\bibfnamefont {M.}~\bibnamefont {Stock}},\ }\href {\doibase
  10.1088/1681-7575/ab860c} {\bibfield  {journal} {\bibinfo  {journal}
  {Metrologia}\ }\textbf {\bibinfo {volume} {57}},\ \bibinfo {pages} {045009}
  (\bibinfo {year} {2020})}\BibitemShut {NoStop}%
\bibitem [{\citenamefont {Alberts}\ \emph {et~al.}(2002)\citenamefont
  {Alberts}, \citenamefont {Johnson}, \citenamefont {Lewis}, \citenamefont
  {Walter}, \citenamefont {Raff},\ and\ \citenamefont {Roberts}}]{avidity}%
  \BibitemOpen
  \bibfield  {author} {\bibinfo {author} {\bibfnamefont {B.}~\bibnamefont
  {Alberts}}, \bibinfo {author} {\bibfnamefont {A.}~\bibnamefont {Johnson}},
  \bibinfo {author} {\bibfnamefont {J.}~\bibnamefont {Lewis}}, \bibinfo
  {author} {\bibfnamefont {P.}~\bibnamefont {Walter}}, \bibinfo {author}
  {\bibfnamefont {M.}~\bibnamefont {Raff}}, \ and\ \bibinfo {author}
  {\bibfnamefont {K.}~\bibnamefont {Roberts}},\ }\href@noop {} {\emph {\bibinfo
  {title} {Molecular Biology of the Cell 4th Edition: International Student
  Edition}}}\ (\bibinfo  {publisher} {Routledge},\ \bibinfo {year}
  {2002})\BibitemShut {NoStop}%
\end{thebibliography}%

\end{document}